\documentclass[10pt,journal,compsoc]{IEEEtran}
\usepackage{graphicx,subfigure,amsmath,url}

\ifCLASSOPTIONcompsoc
  \usepackage[nocompress]{cite}
\else
  \usepackage{cite}
\fi
\ifCLASSINFOpdf
\else
\fi
\begin{document}
\title{Blockchain Driven Privacy Preserving Contact Tracing Framework in Pandemics}

\author{Xiao~Li,
        Weili~Wu,~\IEEEmembership{Senior Member}
        and~Tiantian~Chen% <-this % stops a space
\IEEEcompsocitemizethanks{\IEEEcompsocthanksitem Corresponding Author: Xiao Li, Email: xiao.li@utdallas.edu\protect\\
\IEEEcompsocthanksitem This work was supported in part by the US National Science Foundation (NSF) under Grant NO.1822985.}% <-this % stops an unwanted space
\thanks{Manuscript received April 19, 2005; revised August 26, 2015.}}


\markboth{Journal of \LaTeX\ Class Files,~Vol.~14, No.~8, August~2015}%
{Shell \MakeLowercase{\textit{et al.}}: Bare Demo of IEEEtran.cls for Computer Society Journals}



\IEEEtitleabstractindextext{%
\begin{abstract}
Contact tracing has been proven an effective approach to control the virus spread in pandemics like COVID-19 pandemic. As an emerging powerful decentralized technique, blockchain has been explored to ensure data privacy and security in contact tracing processes. However, existing works are mostly high-level designs with no sufficient demonstration and treat blockchain as separate storage system assisting third-party central servers, ignoring the importance and capability of consensus mechanism and incentive mechanism. 
In this paper, we propose a light-weight and fully third-party free \textbf{B}lockchain-\textbf{D}riven \textbf{C}ontact \textbf{T}racing framework (BDCT) to bridge the gap. In the BDCT framework, RSA encryption based transaction verification method (RSA-TVM) is proposed to ensure contact tracing correctness, which can achieve more than 96\% contact cases recording accuracy even each person has 60\% probability of failing to verify the contact information. Reputation Corrected Delegated Proof of Stake (RC-DPoS) consensus mechanism is proposed together with the incentive mechanism, which can ensure timeliness of reporting contact cases and keep blockchain decentralized. A novel contact tracing simulation environment is created, which considers three different contact scenarios based on population density. The simulation results demonstrate the effectiveness, robustness and attack resistance of RSA-TVM and RC-DPoS in the proposed BDCT. 
\end{abstract}

}

\maketitle

\IEEEdisplaynontitleabstractindextext

% \fi
\IEEEpeerreviewmaketitle

\IEEEraisesectionheading{\section{Introduction}\label{sec:introduction}}

\IEEEPARstart{S}{ince} the first case of the novel corona-virus COVID-19 discovered in December 2019, there have been over 510 million globally confirmed cases, including 6 million deaths by April 2022\footnote{https://covid19.who.int/}. 
The COVID-19 pandemic has brought considerable degree of fear, emotional stress and anxiety among individuals around the world~\cite{Dong2020AnIW}. 
The virus causes severe acute respiratory infection, bringing symptoms such as cough, fever, fatigue and breathlessness, which are very similar to symptoms caused by regular influenza. In addition, the high contagiousness makes it even hard to be controlled. 
In order to help people who have contact with the patient get medical treatment timely, it is imperative to record the contact histories of the patients.

 The World Health Organization (WHO) has announced the importance of contract tracing since EBOLA outbreak in 2014~\cite{world2014contact}. Formally, the contact tracing is the process of identifying history contact cases of people who may have come into contact with infected patients. Many countries have developed contact tracing methods, such as Trace Together in Singapore~\cite{Bay2020BlueTraceAP} and the QR code System in China~\cite{Liang2020}. Some technology companies also developed contact tracing tools, such as Google and Apple developed a Bluetooth-based API that can be used by third parties to develop smart phone apps~\cite{apple_newsroom_2021}. These apps mostly use Bluetooth to recognize nearby devices or GPS signal to get the accurate location coordinates to determine contact cases. Most of these tracing systems rely on central servers controlled by governments or healthcare authorities, which may collect the users' identities and other privacy data through an application installed on smart phones.

Systems based on centralized servers suffer single-point failure and are weak to attacks. Decentralized contact tracing methods are then promoted, which give more control to users. In decentralized model, users are not required to update all data to the server. They can hold data locally, and share their data when necessary.

 As an emerging decentralized data generating, sharing and storing technique, blockchain systems are introduced to solve contact tracing tasks to promote the security and privacy. Blockchain stores data into blocks that are connected to each other as a chain. The data stored in blocks are not able to be tempered. Smart contract deployed on blockchain can perform various functionalities. Furthermore, encryption and anonymization technologies can be applied in blockchain system to protect user's identity. The consensus mechanism in Blockchain allows blockchain systems keep working stably without a central server. 
 
There are some initial attempts of contact tracing systems using blockchain technologies. Hasan et al.~\cite{hasan2021covid} propose proof of location and develop smart contracts to ensure the privacy of contact list. However, no simulation is provided. In addition, there is no incentive mechanism to motivate users to join the system. Authors assume there are plenty of users in the system behaving honestly, while the situation is hard to achieve in practice.
Xu et al.~\cite{DBLP:journals/iotj/XuZOFBI21} proposes BeepTrace blockchain-based contact tracing solution, where a blockchain system plays the neutral role in bridging data transmission between different parties, such as patients, doctors and government authorities. The users' geodata are securely preserved in specially designed blockchain. However, the efficiency of this system is not demonstrated, and no specific consensus mechanism and incentive mechanism are specified in the paper.  Lv et al.~\cite{DBLP:journals/tnse/LvWJCQZ22} proposes Bychain, a three-layer contact tracing framework without reliance on trusted third parties. Proof of Location (PoL) is proposed to verify the contact record and incentive mechanism is design for maximizing contact tracing range. However, Bychain is not able to produce person-to-person accurate contact cases. 

We conclude 4 main challenges to develop a third-party free blockchain-based contact tracing method. The 4 challenges are overlapped with each other: 1) Instead of simply treating blockchain as a separated storage method, how to leverage powerful consensus mechanism in blockchain system to promote data security; 2) How to design an effective consensus mechanism to organize data storage and meanwhile achieve low latency of recording contact information. The popular consensus mechanisms are usually too computational expensive for mobile devices, and may bring significant delay of recording contact information. People should be able to access latest contact records timely to prevent further possible virus spread. 3)  How to design the incentive mechanism so that people are motivated to join the contact tracing system and behaviour honestly.  
4) Due to lack of real-world contact data, as well as high cost of testing whole system in practice, it is hard to evaluate the effectiveness and efficiency of whole systems. % People are working or living in various areas in different ways. For example, some people may living in city where contact can easily happen, while some people living in rural areas are not usually gathering around. Some people's job may requires meeting people face to face every day, while some people can work remotely at home to avoid contact. 
The difficulty of collecting real-world contact information, is not only from privacy concerns, but also diversity of people contact scenarios. Contact cases happened in crowded cities and those happened in rural areas are totally different scenarios with different frequencies and amounts. This diversity brings challenge to design incentive mechanism fair to every one.

There is seldom work that clearly addresses all above mentioned 4 challenges. 
In addition, according to the survey conducted at multiple countries in~\cite{Altmann2020AcceptabilityOA}, though most people accept app-based tracing methods, the concern about the security and privacy is still an obstacle to the common adoption of tracing apps in many countries. Therefore, in this article, we aim to tackle the 4 challenges with the users' privacy ensured by proposing a fully third-party free contact tracing framework with blockchain technology. A RSA based transaction verification algorithm is proposed to ensure the correctness of recorded contact information and improve system robustness. To efficiently store contact information into blocks, we propose Reputation Corrected Delegated Proof of Stake (RC-DPoS) consensus mechanism, which can control the right of appending new blocks. An incentive mechanism is then designed to work with RC-DPoS motivating people to work honestly and maintaining system decentrality.
Finally, we design a contact tracing simulation method that simulates different real-world people contact scenarios to evaluate the effectiveness of proposed framework. 

The reminder of this article is organized as follows.  In Section~\ref{sec:rel_work}, we discuss existing related work on contract tracing. Section~\ref{sec:con_tra_ov} is dedicated to presenting the overview of proposed contact tracing framework. Next we elaborate transaction verification algorithm, RC-DPoS and incentive mechanism in Section~\ref{sec:Tran_Veri}, Section~\ref{sec:rcdpos} and Section~\ref{sec:inc_mec}, respectively. Experimental simulation and discussion are conducted in Section~\ref{sec:simu}. Finally, Section~\ref{sec:con} concludes the paper.

\section{Related Work}
\label{sec:rel_work}
\subsection{Traditional Contact Tracing Methods}
Contact tracing refers to the process that records the people contact history so that the people contacted with a patient can be informed and get medical treatment timely to control the spread of virus.

Various contact tracing tools have been developed using location technologies such as GPS, Wifi, cell tower signal and Bluetooth~\cite{Altuwaiyan2018EPICEP,Prasad2017ENACTEA,nisar2020privacy,Trivedi2021WiFiTraceNC,Hekmati2021CONTAINPC,Ng2021COVID19AY}. 

Reichert et al.~\cite{reichert2020privacy}  propose a
centralized contact tracing method, which assumes every user has their location history stored in their devices, and the health authorities are able to read the data. 
Nisar et al.~\cite{nisar2020privacy} propose to use call data record to trace the patient once she/he is diagnosed positive. However, the trace build is not practical since most people don not answer phones calls very frequently during a day, therefore we can only get limited number of locations.

The contact tracing system based on GPS signal are not reliable for in-door situations, while in-door contact is one major way of virus spread due to short contact distance and long contact time.
Some work proposes to use WiFi or wireless access points to discover contact cases~\cite{Trivedi2021WiFiTraceNC,Altuwaiyan2018EPICEP,Prasad2017ENACTEA}. These frameworks require users to connect their devices to specific wireless access points. However, in some public areas such as shopping malls, airports and train stations, people may not join public WiFi due to network security concerns.  Bluetooth technology can scan nearby devices and get device identities within a small range, which can help generate the contact cases~\cite{Ng2021COVID19AY}. In this article, we also leverage this advantage to protect users privacy that avoids disclosing users' real identities and specific locations. 

Chan et al.~\cite{chan2020pact} propose PACT protocol, where every user holds the contact tracing data on their own local devices, and only when they are tested positive, they will broadcast their contact information to a public platform.  Every other user will check the list on the platform to confirm if they have contacted with anyone in that list. Though this protocol is a third-party-free mobile contracting protocol and easy to be implemented in practice,
However The users are not guaranteed to behave honestly, and the public platform is easy to be compromised since it is open to anyone. 

Most of existing works are centralized where third-party servers are used collecting user's personal data and contact history to match contact records~\cite{DBLP:journals/iotj/AzadAARAIA21}. Centralized models are exposed to risks of single point failure, privacy data leaking and security compromising. Though some contact tracing methods are proposed to be decentralized~\cite{DBLP:journals/access/AhmedMXRMKSHJJ20,chan2020pact}, these methods still require a server to process data computing functions and are vulnerable to dishonest behaviours from malicious users. 

\subsection{Blockchain Based Contact Tracing Methods}
Blockchain technology is first proposed by~\cite{Nakamoto2009BitcoinA} as a distributed ledger for Bitcoin system, which ensures data security without any trust given to third parties. A blockchain system usually constructs a peer-to-peer network, where each user plays exactly the same role and follows the same protocol. Every user stores a whole copy of blockchain, so that all the data on the blockchain are extremely hard to be tempered and single-point failure can naturally be avoided. There's no need of a central server to perform functions in the system, such as collecting, computing or storing data. Users (peers) in a blockchain system have equal rights to perform functionalities by executing smart contracts deployed in the system. A consensus mechanism is enforced in the system to control which user is qualified to generate a new block at each step. An incentive mechanism is also important in the system to motivate users to compete for the right of generating a block. 

Blockchain technology, first known as distributed ledger~\cite{Nakamoto2009BitcoinA}, can make a system work stably without any trust built among parties. With anonymity techniques and data encryption techniques, users in a blockchain system can share data securely without compromising privacy. Blockchain technology has demonstrated significant feasibility in IoT applications, which have similar requirements as contact tracing systems~\cite{DBLP:journals/iotj/GuoDW21,DBLP:journals/tcss/FanWWD21,DBLP:journals/tnse/DingGLW21,DBLP:journals/tr/DongWGSZD21,DBLP:conf/birthday/LuoX0W20}.  

 Blockchain technology shows great potential for developing privacy preserving and efficient contact tracing applications. Idrees et al.~\cite{idrees2021blockchain} point out several challenges and risks associated with the available contact tracing apps and analyze how the adoption of a blockchain-based decentralized network could provide users with privacy-preserving contact tracing.

Besides the BeepTrace~\cite{DBLP:journals/iotj/XuZOFBI21} mentioned above, there exists many other blockchain-based contract tracing frameworks or systems. Arifeen et al.~\cite{arifeen2020blockchain} propose a high-level blockchain based contract tracing framework where blockchain is used for patients to publish contact list. Zhang et al.~\cite{DBLP:journals/csi/ZhangXSZ21} propose PTBM leveraging both permissionless and permissioned blockchain to manage users' location data, and 5G technique provides support for low latency communication. In PTBM, authorized third parties, such as medical centers and medical organizations, are able to compute the contact history and publish history route of patients. 

Peng et al.~\cite{DBLP:conf/sigmod/PengXWHXC21} propose $P^2B$, where users can upload contact information to blockchain storage to be further verified and cross-checked by clients and authorities. $P^2B$ is demonstrated with higher data transmission efficiency than BeepTrace. 
Vangipuram et al.~\cite{DBLP:journals/sncs/VangipuramMK21} propose a three-tier architecture for storing numerous data collected by Internet-of-MedicalThings (IoMT) for contact tracing. In the architecture blockchain is employed to securely transfer the data from the infected person to the hospital system using the edge infrastructure. 

Zuhair et al.~\cite{zuhair2021blocov6} consider a sixth-generation (6G)-assisted unmanned aerial vehicles (UAVs) en-powered mass surveillance system in dense areas, which can monitor body temperature of persons with thermal imaging sensors. Blockchain also works as storage system in their work, and with the powerful bandwidth of 6G, the data can be processed with low latency.
Salimibeni et al.~\cite{DBLP:journals/corr/abs-2108-08275} consider in-door contact tracing scenarios, and propose TB-ICT contact tracing framework,where dynamic Proof of Work (dPoW) credit-based consensus algorithm coupled with Randomized Hash Window (W-Hash) and dynamic Proof of Credit (dPoC) mechanisms are proposed to differentiate between honest and dishonest nodes.
TB-ICT can motivate people to behave honestly since better credit can decrease mining difficulty. However, PoW-based consensus mechanism may bring high computation overhead while BLE-carried devices adopted in the system are not usually computational powerful.

\section{Contract Tracing Framework Overview}
\label{sec:con_tra_ov}
%In this section, we describe overall process of our proposed contract tracing framework. 
\subsection{Problem Definition and Preliminary Settings}
In this paper, we study the contact tracing problem as: given Bluetooth signals on smart devices, with the constraints of preserving privacy, we aim to output pairwise users' contact lists by discovering nearby Bluetooth devices. The goals to achieve for the contact tracing problem are the completeness and correctness of contact list, the contact tracing robustness and attack resistance. 

We assume our \textbf{B}lockchain-\textbf{D}riven \textbf{C}ontact \textbf{T}racing framework (BDCT) is implemented and deployed through clients on smart devices. People can join the contact tracing system by installing the client on their smart devices. It is assumed each user carries one device with the client installed. The client will generate private-public key pair and a unique device ID for each device. The client on a device will use Bluetooth to share the device ID as well as getting device IDs of other nearby devices. Bluetooth is capable to evaluate the distance between two devices within a certain range by the strength of Bluetooth signal. Therefore, the contact distance can be easily computed~\cite{kotanen2003experiments}. The furthest contact distance considered in this paper is 5 meters where Bluetooth can produce strong enough signal to support accurate computation. In this paper, since we directly record the device IDs of contacts rather than record accurate GPS coordinates and match contact information afterwards, without any accurate location data recorded, privacy will be preserved.

\subsection{Contact Tracing Procedure}

\begin{figure*}[htb!]
    \centering
    \includegraphics[width = 0.8\textwidth]{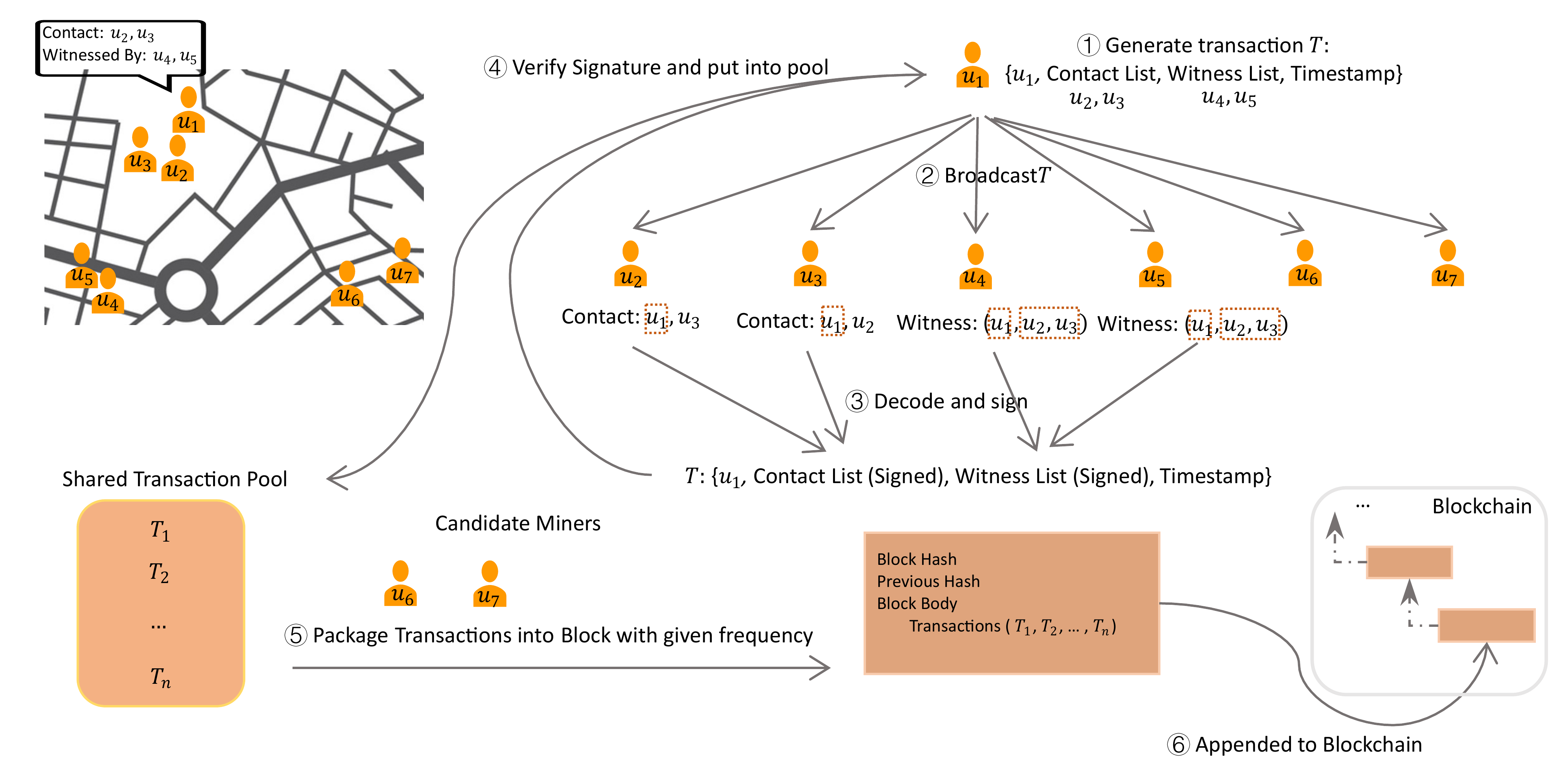}
    \caption{Overview of Contact Tracing Framework (User $u_1$ generates a transaction))}
    \label{fig:tracingfw}
\end{figure*}

At a given frequency, the client on a smart device will scan and record all the device IDs of nearby devices within a range. This process is fast and secure since the client only scans surrounding devices without having to establish stable connection to them, which also avoids cyberattack through Bluetooth channel. If there is a device detected within 2 meters\footnote{The distance can be adjusted according to particular scenarios.}, the client will identify this as a contact case.  The client will then store the device IDs of contacted devices into contact list locally in a special format which will be specified in next section.

Most of previous works ignore the fact that mobile devices are not as robust as computers in terms of internet connectivity, system robustness and security level. The device may fail to collect the contacted device information, or be attacked to  record false contact list. %In our framework, we simulate practical human activities, that people usually contact to others in public area where there's people exists in a wide range.       
To improve the data integrity, a special role \emph{\textbf{witness}} is proposed in this paper.
All the devices that are 2 meters away but still within 5 meters from the current device are considered witnesses of the contact case. Witnesses play important roles in BDCT, which help verify the reported contact list, speed up the verification process, and recover the missed contacts.  The client will also store the device IDs of witnessed devices into witness list locally in the similar format as contact list.  With the pseudo IDs of contact devices and witness devices stored locally, users can check whom they have contacted with and who have witnessed their contact cases at any time without knowing the real identity of the device owner. 

Based on the above setting, we now illustrate the whole Blockchain-Driven Contact Tracing Framework (BDCT) in Figure~\ref{fig:tracingfw} with an example.

In Figure~\ref{fig:tracingfw}, at a given \emph{Timestamp}, assume user $u_1$ would like to report his current contact case. He will initiate a contact tracing transaction $T_{con}$. let's assume users $u_2$ and $u_3$ are within 2 meters from $u_1$ and hence considered a contact case with user $u_1$. Users $u_4$ and $u_5$ are 2 meters away but still within 5 meters from $u_1$, and they witness that $u_1$ is with $u_2$ and $u_3$. $u_6$ and $u_7$ are considered irrelevant to this contact case. As in the figure, there are 6 steps from generating contact record as blockchain transaction to the transaction being stored to blockchain storage in every device. 

\emph{\textbf{Step 1}}: User $u_1$ initiates a blockchain transaction $T_{con}=\{T_{id}, u_1, Contact List, Witness List, Timestamp\}$, which is used for record the contact case of $u_1$ at $Timestamp$. One transaction represents one contact case of users at some timestamp. $T_{id}$ is an unique transaction ID for each transaction. $Timestamp$ is the exact time that $u_1$ contacts the users in $Contact List$. The $Contact List$ and  $Witness List$ contains secret messages from $u_1$ encrypted by the public keys of each contacted devices ($u_2$ and $u_3$) or witness devices ($u_4$ and $u_5$). Formal definition of $Contact List$ and  $Witness List$ will be presented in Section~\ref{sec:Tran_Veri}. 

\emph{\textbf{Step 2}}: User $u_1$ then broadcasts the transaction $T_{con}$ through internet to every user who have the client installed. Since no one knows others' identities, $u_1$ is not able to directly send message to $u_2$,  $u_3$,  $u_4$ and  $u_5$. 

\emph{\textbf{Step 3}}: When other users receive the transaction $T_{con}$, it will check if it contacted with $u_1$ at $Timestamp$ or if it witnessed the reported contact case. Then the contacted users in this example, $u_2$ and $u_3$, and the witnessed users, $u_4$ and $u_5$, will try decode the received message, sign the decoded message and broadcast this signed transaction. The transaction generator $u_1$ will receive the signed transaction. 

\emph{\textbf{Step 4}}: After receiving the signed transaction $T_{con}$,  $u_1$ will verify the signature by decoding the signature with contact's or witness's public key to make sure the contact list and witness list are signed by correct people. $u_1$ will wait for the signatures within a specific delay $d$, such as 60 minutes. Only the records in $Contact List$ verified valid will be finally preserved in $T_{con}$. 
%The full list of $Contact List$ is considered valid if all the tuples in $Contact List$ are signed by correct persons, or at least one witness confirm that $Contact List$ is valid by signing the corresponding tuple in $Witness List$. If still not all contact records are confirmed within $d$, $u_1$ will only preserve the signed ones, and put filtered 
$u_1$ will put transaction $T_{con}$ into a shared transaction pool which is synchronized on every device along with blockchain. 

\emph{\textbf{Step 5}}: At given frequency, one of the candidate miners will be selected to package all the transactions in the transaction pool into a block. In this paper, we propose the  Reputation-Corrected DPoS (RC-DPoS) mechanism to choose the candidate miners, which will be presented in detail in Section~\ref{sec:rcdpos}. 

\emph{\textbf{Step 6}}: The block is finally appended to the blockchain by the miner, and broadcast to all users in the network for synchronizing. 

Step 1 to Step 4 will be elaborated in Section~\ref{sec:Tran_Veri} by proposing RSA-Based transaction verification method. In Section~\ref{sec:rcdpos}, RC-DPoS and corresponding incentive mechanism are presented to complete Step 5 and Step 6.

\section{Transaction Verification Method}
\label{sec:Tran_Veri}

In this section, we will first describe how to initialize credentials for each user and then present \emph{\textbf{RSA-based Transaction Verification Method}} (RSA-TVM).
There are two major goals on contact tracing system: 1) data integrity: the collected contact cases should be as complete, untampered and correct as possible; and 2) privacy: the whole system should never initiatively disclose any location or identity information of users.  

In this paper, We propose \emph{\textbf{RSA-based Transaction Verification Method}} (RSA-TVM) to make sure the contact records in the transaction are valid meanwhile ensuring the anonymity.  We employee RSA algorithm as encryption module~\cite{rivest1978method}.  RSA algorithm is an asymmetric encryption algorithm, and is able to generate a key pair, (public key, private key) for a user. Public key is known by public, while the private key is only known by the owner.  The secret message encrypted by public key can only be decoded by private key owner. A message can be signed by private key indicating owner's consent to the message, and the signed message can be verified by corresponding public key to ensure the signature is correctly signed by the private key owner. 

%To implement the verification algorithm, each user must first initial unique credential. 
Next we will first describe how to initialize credentials for each user and then present RSA-TVM.

\subsection{Generate User Credential}

When a person $u$ installs the tracing client on a smart device and become an user of the tracing system, the client will first name the device with an unique device ID, denoted as $u_{D_{ID}}$, and then generate a RSA key pair (public key, private key), denoted as $(u_{Pub\_{key}},u_{Pri\_{key}})$. The length of each key is set to 1024 bits. The private key will be stored locally in the smart device. The public key and the device information will be included in a transaction through the client, then be stored into blockchain. This transaction is called ``\emph{\textbf{Registration Transaction}}'', which is defined as $T_{reg} = \{T_{id}, \{u_{D_{ID}}, u_{Pub\_{key}}, t\}\}$. $T_{id}$ is the unique id for each transaction and is generated by SHA256 algorithm~\cite{DBLP:conf/sacrypt/GilbertH03} based on timestamp $t$ as well as the transaction content $\{u_{D_{ID}}, u_{Pub\_{key}}, t\}$, so that any change made on the content will cause a different $T_{id}$.

 After the registration transaction is stored in the blockchain, since every user in the system have a synchronized copy of the whole blockchain, every user will hold the public keys for every others.
Users are able to modify their device ID or credentials by submitting a new  registration transaction, so that every other users can get a new copy of the updated device ID or public key.

Users will scan the nearby devices (through Bluetooth) at a given frequency to get the nearby devices' IDs and record them locally. We avoid any device connection through Bluetooth channels to improve security. The client only collects the devices' IDs, and look up the registration transactions to get the public keys for generating secret message later used in $ContactList$ or $WitnessList$. 
Next if the user wants to report contact cases and store the contact information into Blockchain, ``\emph{\textbf{Contact Transaction}}'' will be initialized.

\subsection{RSA-based Transaction Verification Method (RSA-TVM)}
\label{sec:RSATVM}

\begin{figure*}[htb!]
    \centering
    \includegraphics[width = 0.8\textwidth]{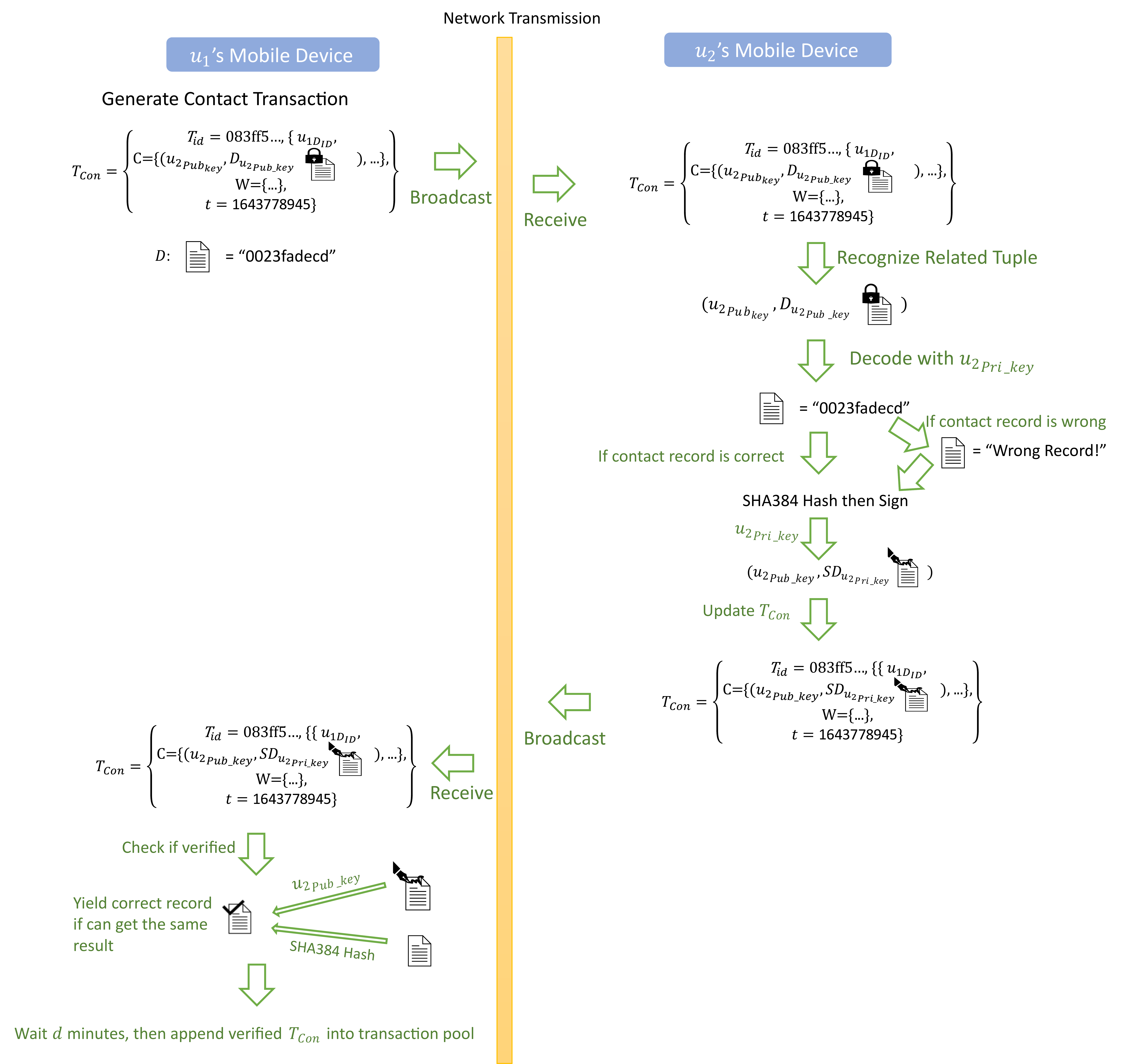}
    \caption{An Example of RSA-TVM}
    \label{fig:TVM}
\end{figure*}

If there are nearby devices within 2 meters detected by user $u$'s device, $u$ can generate  ``\emph{\textbf{Contact Transaction}}'', denoted as $T_{con}=\{ T_{id},\{ u_{D_{ID}}, C, W, t\}\}$, where $T_{id}$ is the unique id for each transaction and is generated based on timestamp $t$ and the transaction content $\{ u_{D_{ID}}, C, W, t\}$. The $u_{D_{ID}}$ is the device ID of $u$, and $t$ is the timestamp for this contact case. $C$ and $W$ are Contact List and Witness List,  which contain the information of the contacted people(devices) and the witness of this contact case, respectively. To generate $C$ and $W$, user $u$ first needs to decide a original secret message $D$, and then encrypt it with the public key of the contacted people (e.g., $u_{i}$) and the witnesses (e.g., $u_{j}$) of this contact case. For each contacted person $u_{i}$, ${u_i}_{Pub\_key}$ encrypted text, denoted as $D_{{u_i}_{Pub\_key}}$ is generated.  Similarly, for each witness $u_j$, $D_{{u_j}_{Pub\_key}}$ is generated. Formally, the Contact List $C$ is defined as a set of tuples: $C = \{ ({u_i}_{Pub\_key},D_{{u_i}_{Pub\_key}}) | \forall u_i\}$. Similarly, the Witness List is defined as: $W = \{ ({u_j}_{Pub\_key},D_{{u_j}_{Pub\_key}}) |\forall u_j\}$. 

Ideally, The secret message should be unique for every $u_i$ and $u_j$ for every transaction to ensure security which requires secret message as long as possible. However, the encryption time increase rapidly with the length of text. In practice, we set each secret message contains 10 Hex characters ($0-9,a-f$), which is able to represent about $1.1\times 10^{12}$ different messages. 

Witness list $W$ can be very helpful to avoid contact case loss and improve robustness against dishonest user behaviors or system failure. We will show this later in Section~\ref{sec:simu}.

The transaction $T_{con}$ will then be broadcast to all users in order to protect privacy. Each user will check if $C$ or $W$ in the received $T_{con}$ contains his/her public key. If so, the related tuples require his/her verification. Since the messages are all encrypted, therefore only the user who holds the public key can decrypt the encrypted secret message by his/her public key.

When $u_i$ identifies the tuple $({u_i}_{Pub\_key},D_{{u_i}_{Pub\_key}})$ in $C$, $u_i$ will decrypt the encrypted text $D_{{u_i}_{Pub\_key}}$ with the private key ${u_i}_{Pri\_key}$ to get the secret message $D$. Then $u_i$ will check local contact history. If $u_i$ has the record that $u_i$ contacted with $u$ at timestamp $t\pm 3\ min$, then $u_i$ can confirm the tuple $({u_i}_{Pub\_key},D_{{u_i}_{Pub\_key}})$ valid in $T_{con}$. 
%To let the transaction generator $u$ confirm that the contact list is verified by correct person, $u_i$ signs the secret message $D$ with private key ${u_i}_{Pri\_key}$. The signed $D$ can be verified by
Then $u_i$ needs to send a message back to $u$ to indicate that the contact record about $u_i$ in $T_{con}$ is confirmed. Specifically, $u_i$ signs the secret message $D$ with his private key. The signed text is denoted as $SD_{{u_i}_{Pri\_key}}$. Then $u_i$ replaces $({u_i}_{Pub\_key},D_{{u_i}_{Pub\_key}})$ with  $({u_i}_{Pub\_key}, SD_{{u_i}_{Pri\_key}})$ in $T_{con}$, and broadcast to all users.

$u_j$ will conduct similar verification on the related tuple in witness list $W$. If $u_j$ has the record that the transaction generator $u$ contacted with all users in $C$ at $t$, then $u_j$ will consider all tuples in $C$ valid by signing secret message in related tuple in $W$. 

If $u_i$ can not find any local record showing $u_i$ contacted with $u$ at timestamp $t\pm 3\ min$, then $u_i$ believes this is a wrong record. $u_i$ will sign a predefined warning message $Z$=\emph{\textbf{"Wrong Record"}} instead of signing the secret message $D$. The tuple $({u_i}_{Pub\_key},D_{{u_i}_{Pub\_key}})$ in $T_{con}$ will then be $({u_i}_{Pub\_key}, Z_{{u_i}_{Pri\_key}})$. Once the transaction generator $u$ receives the updated $T_{con}$ from user $u_{i}$, $u$ will verify the signature with the public key of $u_{i}$.

A tuple in contact list $C$ in $T_{con}$ is considered valid if: 1) there is no signed warning message in the tuple, and 2) the secret message in tuple is correctly signed by the contacted person or at least one tuple in witness list is correctly signed by the witness. Due to network or system failure of users' smart devices, users may have no response to the related tuple in $C$ or $W$ within given delay $d$. In this case, the tuple will still be considered valid as long as one witness has verified the this contact case is correct. 

If not all tuples in the contact list $C$ are verified valid within a specific delay $d$, then only the valid tuples in $C$ will be preserved in $T_{con}$. The transaction $T_{con}$ will be put into the shared transaction pool waiting to be mined, e.g. permanently stored in blockchain. Any user $u_i$ or $u_j$ who signed $D$ or $Z$ will get reward for helping verify the contact case. We will discuss reward policies in Section~\ref{sec:inc_mec}.
Figure~\ref{fig:TVM} shows the process of RSA-TVM that $u_2$ verifies $u_1$'s contact case.

\section{ Reputation-Corrected DPoS mechanism (RC-DPoS)}
\label{sec:rcdpos}

Most existing work considers it is straightforward to let the transaction generator directly package the verified transactions into blocks and then broadcast to all users instead of choosing a miner to do the job. However, the above strategy will cause \emph{\textbf{unfair incentive reward problem}} due to the nature of diverse contact scenarios. 

\emph{\textbf{unfair incentive reward problem}}:  Users are rewarded for reporting contact cases by generating contact transactions. However, people have different chance to have contact cases due to the diversity of jobs or living styles. People who live or work in human-dense areas, such as cashiers in markets and staffs at transport stations, will obviously have much more contact cases than those who stay or work at home, thus gain much more reward. This will even encourage people to go out and make contacts in order to earn reward, which is against the social distancing policy during pandemics. 

 In addition, since miners can be the one not in the $Contactlist$ or $WitnessList$, it helps avoid group cheating that small groups deliberately generate fake contact cases, verify contact transactions for each other, package transactions and append new blocks in order to gain great amount of reward rapidly. Therefore, it is imperative to carefully design consensus mechanism and incentive mechanism to balance the reward. The consensus mechanism is required computational lightweight and have high transaction throughput to satisfy the huge data storage demand on smart devices which are usually have low computational power. 
 
 %The miners are selected based on users' credit and current balance in next section by proposed Reputation-Corrected DPoS mechanism.

The Delegated Proof of Stake (DPoS) consensus mechanism~\cite{larimer2014delegated} is a popular light-weight consensus mechanism. DPoS provides high-speed consensus making so that emerging transactions can be stored into blocks timely. In DPoS consensus mechanism, each user holds some stakes, which are usually crypto-currency. Whenever there is no candidate miners, every user will vote someone they trust. The weight of the vote is proportional to the stake of the voter. That is, more stake gives the voter more vote power. After the voting, the users with top $k$ total weighted votes will be selected as $k$ candidate miners. Whenever there is block waiting to be appended into blockchain, one candidate miner will be randomly chosen to do the job, and the chosen miner will be removed from the candidate miner set once the job is done. Once candidate miner set is empty, new round of voting will start. 

DPoS can produce high throughput without compromising decentrality of blockchain system if everyone is honest and the voting is random. However, it can not be directly applied in our proposed BDCT. In order to motivate people to share their contact information, reward must be given to those who generates contact transactions honestly. In DPoS the reward is stake, people living or working in human-dense areas will gather stakes quickly. Thus their votes will gradually become highly weighted due to high stakes, hence their votes will easily determine the selection of candidate miners. In other words, the whole blockchain system will be dominated by those people who generate contact cases often. 

In order to solve the issue described above, we propose Reputation-Corrected DPoS (RC-DPoS) consensus mechanism. In RC-DPoS, we assign reputation to each user, which is represented by credit $c$. Users will gain reputation reward instead of stake reward for honestly reporting their contact cases, while only gain stake reward for working as a miner. Specifically, the RC-DPoS mechanism works as follows:

\emph{\textbf{Step 1}}: When new users first join the contact tracing framework, they will be initialized with a fixed start-up stake $s_0$ and credit $c_0$.

\emph{\textbf{Step 2}}: Initially, the candidate miner set is empty, the candidate selection process will start. Each user votes for another one trusted user and users can not vote for themselves. Similar to DPoS, the vote is weighted according to the voter's stake. But the total votes received by a user will be corrected by receiver's credit. Formally, let $N$ denotes the total number of users in the system. For user $u_i, i \in Z^N$, the total vote score accumulated by $u_i$ is calculated according to Equation~\ref{eq:g}:
\begin{equation}
    \label{eq:g}
    G_i = \frac{ RF(u_i) +1}{2} \sum_{u_k} \frac{s_k}{\sum_{j \in Z^N} s_j}
\end{equation}
where the sum taken over user $u_k$ who votes $u_i$ is the total weighted vote received by $u_i$, $s_k$ is the current stake amount of $u_k$.  $c_i$ is the current credit amount of $u_i$. $RF(u_i)$ is the \emph{\textbf{reputation correction factor}} of user $u_i$, which is defined as:
\begin{equation}
    RF(u_i) = \frac{c_i- min_{j \in Z^N}(c_j)}{max_{j \in Z^N}(c_j) -min_{j \in Z^N}(c_j)}.
\end{equation}
$RF(u_i) \in [0,1]$, and $\frac{ RF(u_i) +1}{2} \in [0.5,1]$. 
The intuition behind this equation is that users with good reputation should have higher chance to be a candidate miner in order to improve the system security, meanwhile we also avoid applying too much punishment on other users with lower reputation (maximum 50\% off on received votes).  %Another benefit of $G_i$ is that it can balance the probability being chosen as candidate minder from different users that have different contact frequency. 

\emph{\textbf{Step 3}}: Rank all users in descending order according to their vote scores. The top $\lceil N/5 \rceil$ users are selected into candidate miners set. The size of candidate miners set can be adjusted based on specific applications. 

\begin{figure*}
    \centering
    \includegraphics[width=0.8\textwidth]{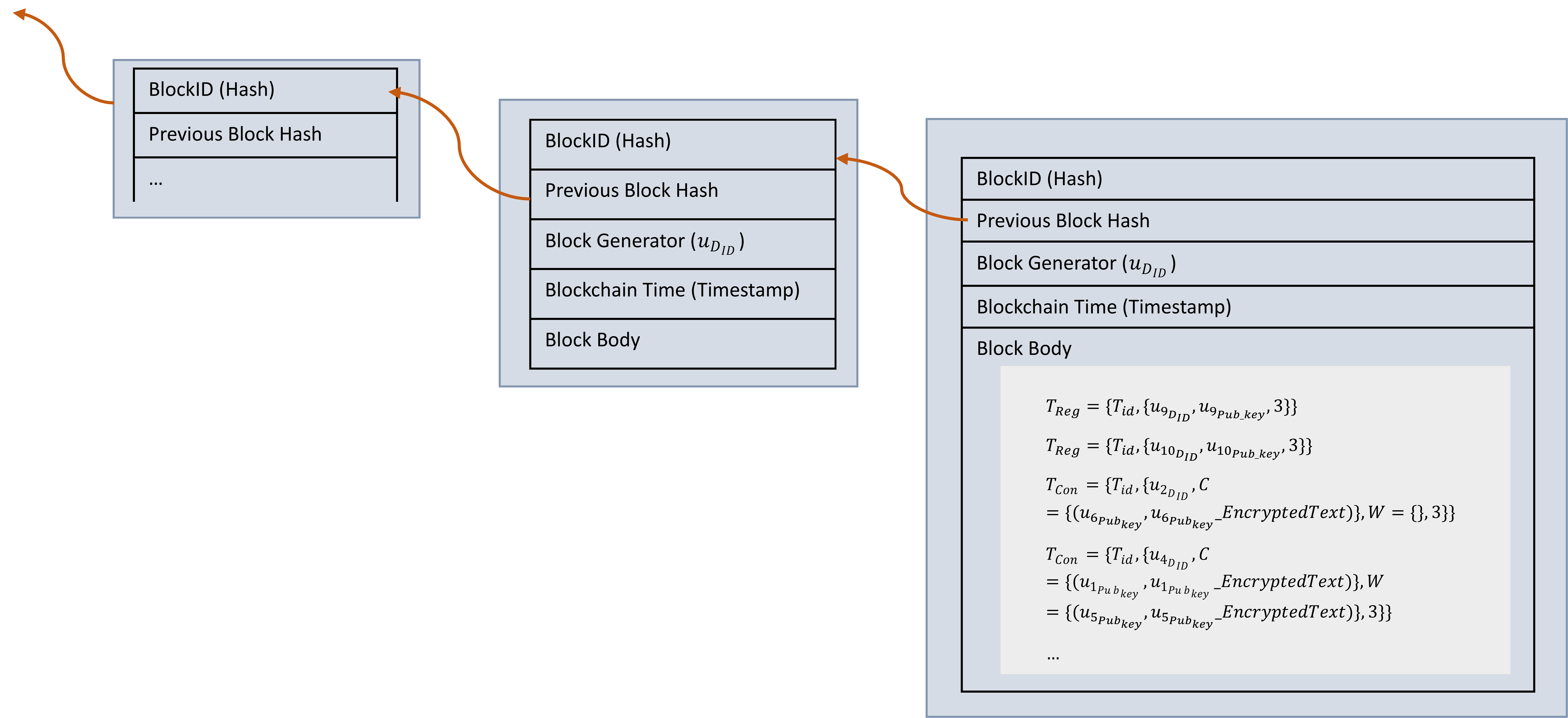}
    \caption{Blockchain Storage Structure in Proposed Framework}
    \label{fig:blockstr}
\end{figure*}

\emph{\textbf{Step 4}}: At a given mining frequency (3 minutes, 5 minutes or so on),  one arbitrary miner selected from the candidate miner set will package all the transactions in the shared transaction pool into a block and append it into the miner's local blockchain. Then shared transaction pool is empty and waits for new verified transactions. The structure of the blockchain storage is illustrated in Figure~\ref{fig:blockstr}.

\emph{\textbf{Step 5}}: The miner then broadcasts this blockchain update to all users. Users in the system will update their local blockchain and the local transaction pool. The miner will be given stake reward and reputation reward. Reward detail will be elaborated in Section~\ref{sec:inc_mec}. Then the miner will be removed from the candidate miner set.

\emph{\textbf{Step 6}}: When a miner fails to do this job within a excusable delay (e.g. 10 minutes) due to network disconnection or system failure, a penalty will be applied on the miner by taking away some credits and no stake reward will be given. The miner will be removed from the candidate miner list and another miner will be delegated to do the job. 

\emph{\textbf{Step 7}}: If the candidate miner set is empty, back to \emph{\textbf{Step 2}}. 

\section{Incentive Mechanism}
\label{sec:inc_mec}

The proposed BDCT contact tracing framework is highly automatic without a central server. BDCT fully relies on people to generate transactions, store contact cases into blocks and maintain decentrality, therefore it is crucial to design an incentive mechanism to motivate people to generate contact transactions and append blocks into blockchain honestly. It is also important to ensure the incentive mechanism does not specially benefit a particular group of people to avoid the system becoming centralized and dominated. If the rewards are taking over by a small specific group of people, others will be discouraged, and the whole system will be barely helpful for contact tracing. In this paper, we design the incentive mechanism as a composition of following 4 incentive policies.

\begin{enumerate}
    \item Users will be rewarded with $1$ unit\footnote{numbers in the incentive mechanism only for indicating the relative amount, they can be of any unit.} credit for generating transactions. The users can not get the reward until the transaction is accepted by the shared transaction pool. This will motivate users to honestly report their contact list. With more credits, according to Equation~\ref{eq:g}, users will be more likely to be selected as candidate miner and thus can get more credit reward as well as stake reward. %Please note that the rewards here can not be too high, because inappropriate reward will make the people who contact people very often gather credits too quick, and hence dominate Candidate Miners.
    \item Users will be rewarded with $1$ unit credit after successfully verifying related tuple in contact transactions. This will motivate users to participate in generating transactions and improve the speed of verifying contact cases. %Again, the rewards here can not be too high either. The reason is the same as above.
    \item Users will be rewarded with $R_i$ unit stake reward and $1$ unit credit reward for mining a block, e.g. append a new block into existing blockchain. $R_i$ is corresponding to the total amount of transactions that $u_i$ generated, which is formally defined as:
    
    \begin{equation}
        R_i = w*\frac{ TF(u_i)+1}{2},
    \end{equation}
    
    \begin{equation}
        TF(u_i) = 1 - \frac{t_i- min_{j \in Z^N}(t_j)}{max_{j \in Z^N}(t_j) -min_{j \in Z^N}(t_j)}.
    \end{equation}
    $w$ is a predefined reward amount (e.g. $5$ units) and $t_i$ is the total number of transaction generated by $u_i$. $TF(u_i) \in [0,1]$ is called the \emph{\textbf{transaction correction factor}}. From the above definition, we could find that the more transaction $u_i$ has generated, the lower stake reward will be given to $u_i$. The intuition behind $R_i$ is that we do not want people who generate much transactions gain much stake reward for completing mining job since they naturally have more chance to become a miner according to incentive policy 1). On the other hand, people who generate less transactions will get more stake reward per mining job they complete. $\frac{ TF(u_i)+1}{2} \in [0.5,1]$ will make a maximum 50\% off on the stake reward.  Therefore $R_i$ can help balance the stake reward among users in different contact scenarios, hence help maintain the vote power distributed. 
    \item A user will be punished if the user fails to complete a mining job. $5$ units credits will be deducted on the miner for this punishment.
\end{enumerate}

The stake reward is usually pecuniary crypto-currency, which can be distributed by government or healthcare authorities. Since this system does not require frequent maintenance and huge computation center, the budget can be saved for pecuniary stake reward. Then with accurate and efficient contact tracing, BDCT will save more money for government by helping control the virus spread. 

%If we jointly consider the incentive mechanism and the proposed RC-DPoS, we can see it can roughly achieve balance for all kinds of users. Users who contact many people per day will generate many transactions, and earn credits, which will increase the reputation factor in $G_i$, however, the will not gather stake rewards too quick, since the stake reward is recalculated based on how much transactions the user has generated. For users who don't contact many people, though they have less possibility to be a miner, they will get higher stake reward for each job they do. 

% \section{Security and privacy analysis}
% \label{sec:sec_an}
%  To be Continued. 

\section{Simulation}
\label{sec:simu}
\subsection{Simulation Method}
\label{sec:sim_method}
Though there are some well-known real-word trajectory datasets indicating real people movements in specific areas ~\cite{DBLP:journals/simpra/JahromiZG016,huang2019grab,didi,DBLP:journals/tsas/Mariescu-Istodor18,DBLP:conf/gis/YuanZZXXSH10,DBLP:conf/sensys/LianZ18},  they are mostly based on the record obtained from mobile vehicles or cell phone calls, the trajectories are not continuous or the number of trajectories are not sufficient to support the simulation in terms of frequency and amount.  Since it is hard to collect real-word trajectory in a wide range due to privacy concerns and diversity of contact scenarios, we conduct experiments on synthetic datasets that simulates different people contact scenarios to demonstrate the effectiveness of the proposed BDCT contact tracing framework.  

We propose to consider three general contact scenarios decided based on population density: \textbf{Low density (Sparse)}, \textbf{Medium density (Medium)}, \textbf{High density (Crowded)}. Each scenario can intuitively represent for one kind of real-world people contacting cases.  ``Sparse'' can represent for the people contacting cases in rural area or residential area. ``Medium'' can represent for the cases in schools, parks or other common public areas. ``Crowded'' represents for contacting cases happening in some very crowded places, such as shopping malls and sports events.

People in 3 scenarios have different frequencies of having contact cases, different numbers of contacted people and witnesses. Therefore the frequency of generating transactions, and the length of contact list and witness list need to be adjusted for simulating 3 scenarios. %Users are also needs to be created for different contact cases to test if the framework will benefit one kind of users more than another kinds. 
To achieve this goal, we specify the settings for the three scenarios as follows: 

\emph{\textbf{Low density (Sparse)}} case: In each transaction, the length of contact list and witness list follow normal distribution $\mathcal{N}(\mu=0, \sigma=2)$ and $\mathcal{N}(\mu=0, \sigma=1)$, respectively. The frequency of generating transaction is $1\ cases/hr$. 

\emph{\textbf{Medium density (Medium)}} case: In each transaction, the length of contact list and witness list follow the normal distribution $\mathcal{N}(\mu=2, \sigma=4)$ and $\mathcal{N}(\mu=2, \sigma=2)$, respectively. The frequency of generating transaction is $3\ cases/hr$. 

\emph{\textbf{High density (Crowded)}} case: In each transaction, the length of contact list and witness list follow the normal distribution $\mathcal{N}(\mu=5, \sigma=2)$ and $\mathcal{N}(\mu=7, \sigma=2)$, respectively. The frequency of generating transaction is $12\ cases/hr$.

We implement the framework with Python 3.7, and all simulations are conducted on a machine with Intel Core i7-8750h 8 cores and 32GB memories. Each user is implemented as a thread of python, and all threads are run simultaneously to simulate real time contact. We randomly generate the contact cases for each user without considering reasonable trajectories for them,  since the trajectories does not affect the evaluation of the effectiveness and efficiency of the whole framework. For the length of contact list or witness list, we only adopt the non-negative number sampled from corresponding normal distributions. 

\subsection{Decentrality Evaluation}
It is crucial to maintain the decentrality of BDCT, so that the voting power is distributed and every user can be equally motivated to keep contributing contact cases honestly.

%We first simulate these three scenarios by creating 600 users in each scenario.
We simulate 200 users for each contact scenario, and hence totally 600 users are in the whole simulation environment. To measure the dencentrality of the system, we draw Lorenz curve, and calculate the Gini coefficient/index of three factors of the 600 users: user \textbf{balance} (cumulative stake reward), user \textbf{credits} (cumulative reputation reward) and the \textbf{total number of mined blocks}. 

Lorenz curve is originally proposed for drawing the cumulative income from different units when they are in the ascending order~\cite{kakwani1977}. The closer the income distribution is to uniform distribution, the closer the corresponding Lorenz curve is to line $y=x$. We extend Lorenz curve in this article to illustrate the decentrality of the proposed RC-DPoS. 

The Gini coefficient or Gini index $Gini$ is a metric for quantitatively measure inequality of a distribution, which can derived from Lorenz curve~\cite{dorfman1979formula}. It is defined as a ratio with values between 0 and 1. Specifically, the numerator is the area between the Lorenz curve of the distribution and the uniform distribution line; the denominator is the area under the uniform distribution line. Hence, $Gini=0$ indicates perfect equality of a distribution, and $Gini=1$ indicates the distribution is total skew to one unit. 

We adopt the DPoS mechanism as the baseline. In the baseline, no credit reward is given to users, and users will get $1$ unit stake reward for generating or verifying transactions and $5$ units stake reward for mining a block. Other settings are kept the same as proposed BDCT. The initial stake and credit of users are set to 100 units. Random voting strategy is adopted for voting the candidate miners.

\begin{figure*}[htb!]
	\centering
	\small	
	%\vspace{0.1in}
	\subfigure[Gini coefficient, Baseline DPoS]{
		\label{}
		\begin{minipage}[t]{0.4\textwidth}
			\centering
			\includegraphics[width=1\linewidth]{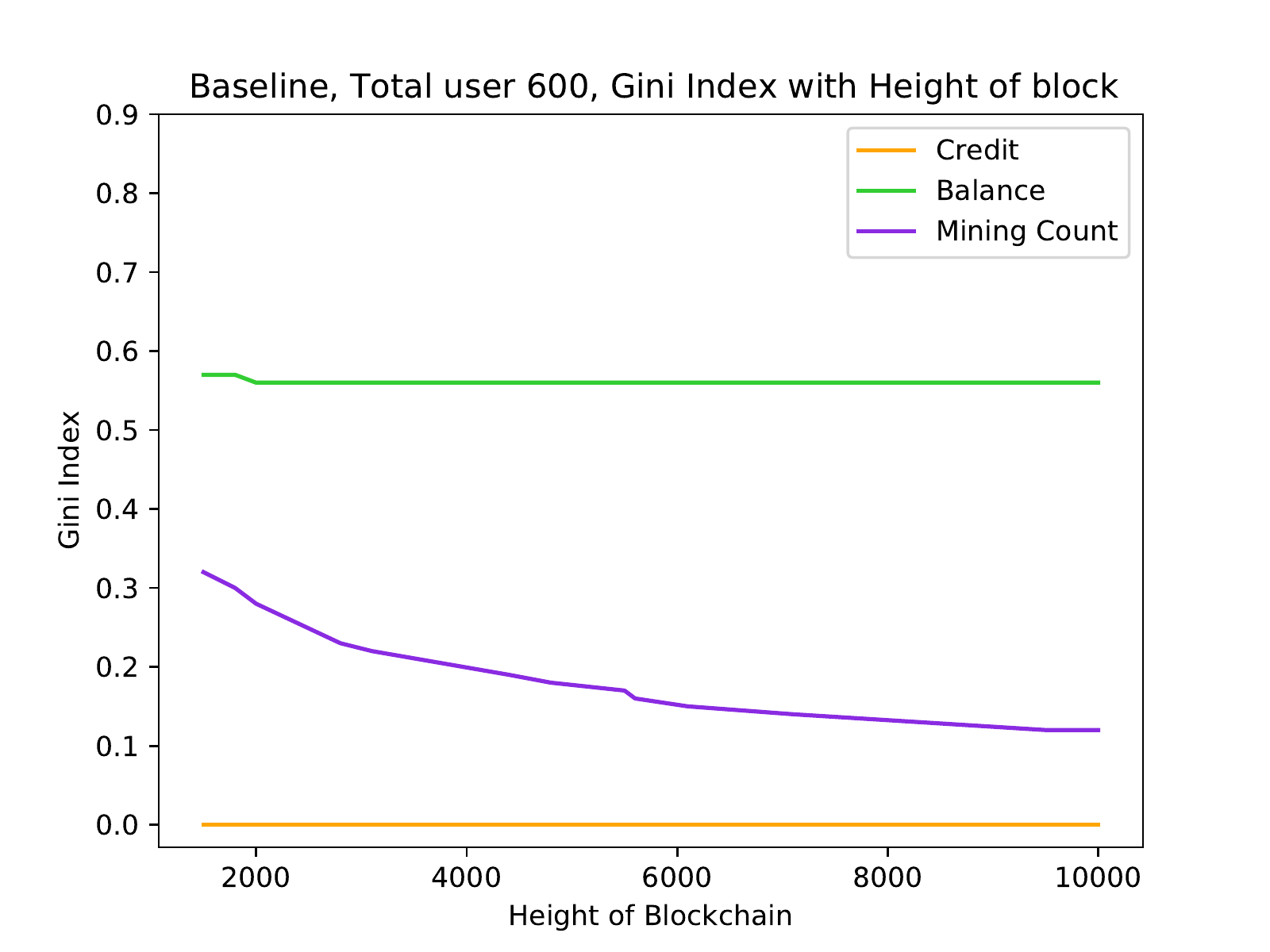}
		\end{minipage}
	}
	\subfigure[Gini coefficient, Our BDCT]{
		\label{}
		\begin{minipage}[t]{0.4\textwidth}
			\centering
			\includegraphics[width=1\linewidth]{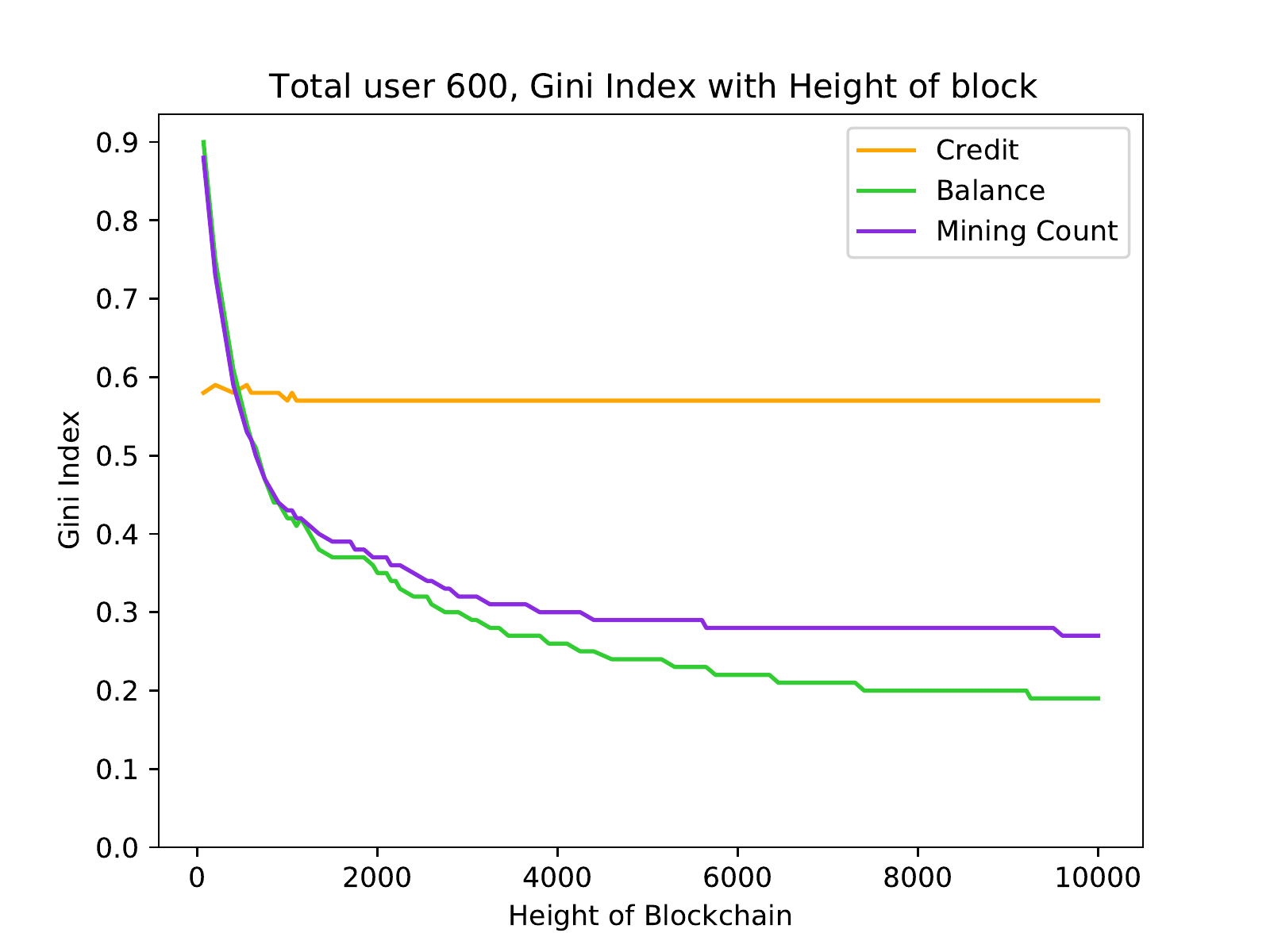}
		\end{minipage}
	}
	\subfigure[Lorenz Curve, Baseline DPoS ]{
		\label{}
		\begin{minipage}[t]{0.4\textwidth}
			\centering
			\includegraphics[width=1\linewidth]{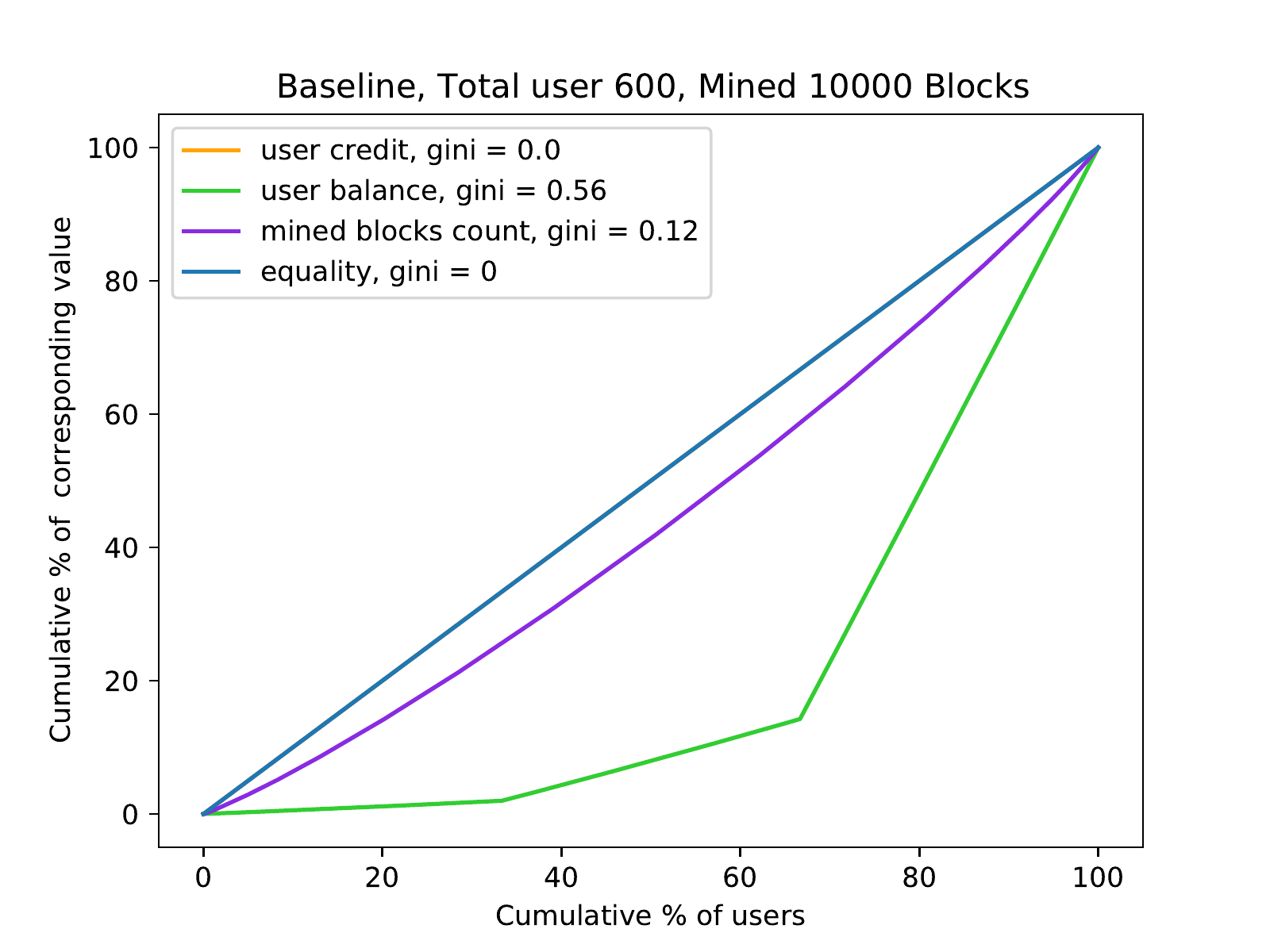}
		\end{minipage}
	}	
	\subfigure[Lorenz Curve, Our BDCT]{
		\label{}
		\begin{minipage}[t]{0.4\textwidth}
			\centering
			\includegraphics[width=1\linewidth]{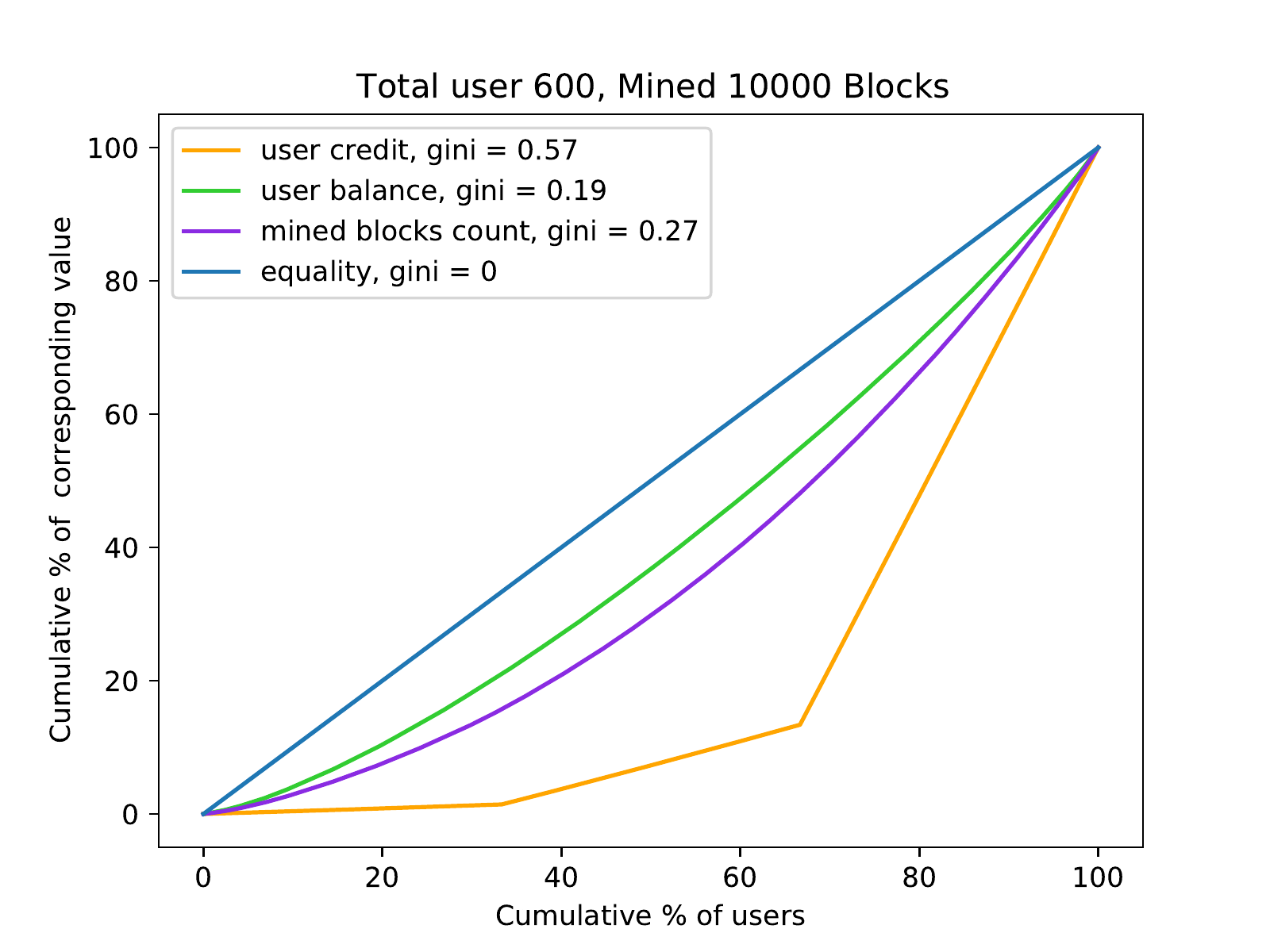}
		\end{minipage}
	}	
	\caption{Lorenz Curve and Gini Coefficient of proposed BDCT framework and baseline DPoS}	
	\label{fig:gini_results}
\end{figure*}

We run the simulation for 10 times, and evaluate statistical significance of Gini coefficient of user balance between baseline and proposed BDCT. We conduct a two-sided T-test for the null hypothesis that two frameworks' stake reward distribution have identical average (expected) values. The $p=1.22\times10^{-32}$ indicates BDCT achieves definitely better stake reward dencentrality. 

Figure~\ref{fig:gini_results} shows the results of Gini coefficient and Lorenz curve of the three factors. Figure~\ref{fig:gini_results}(a) and (b) show how the Gini coefficient change with more and more data stored in the blockchain. Figure~\ref{fig:gini_results}(c) shows the Gini coefficient of balance of baseline DPoS remains as high as 0.56 when blockchain height is 10k, indicating the stake rewards are mostly given to people in Dense scenario. Since the baseline DPoS mechanism does not consider any credit reward, therefore the Gini coefficient of credit is 0 in Figure~\ref{fig:gini_results}(a)(c). In addition The Gini coefficient of mined blocks count of baseline is close to 0, this is because under the random vote strategy in DPoS, users has the same expectation to be selected as a miner. 

Figure~\ref{fig:gini_results}(b) and (d) shows the results of our proposed BDCT framework. 
In Figure~\ref{fig:gini_results}(b), our BDCT framework makes the gini coefficient decrease with the height of blockchain, which means BDCT is achieving balanced stake reward when we continue recording more data. 
In Figure~\ref{fig:gini_results}(d), as expected, the Gini coefficient of credit is 0.57 when blockchain height is 10k, showing users in dense area can indeed earn more credit than other users. The Gini coefficient of mined blocks count is 0.27 which is higher than 0.12 in baseline, indicating people in dense areas indeed have higher chance to be a miner. Gini coefficient of credit is 0.19 which is significantly lower than 0.56 in baseline and demonstrates our RC-DPoS and proposed incentive mechanism can successfully balance the stake reward among different groups of users.

We further investigate the stake reward distribution. In DPoS baseline, the 200 users (1/3 of total users) in the Crowded scenario together hold more than $85\%$ stake rewards giving them more than $85\%$ vote power. On ther other hand, the three groups of users in three different contact scenarios in our proposed BDCT hold stake reward $23\%$, $42\%$, and $34\%$ respectively.

\subsection{Robustness Evaluation}
Mobile devices are usually with low computational power and low security level, and sometimes may suffer from system failure or network delay and disconnection. All those factors can cause failure of detecting contact case, verifying contact list or receiving transactions. 
In this paper, We proposed witness for every contact case in BDCT framework, which can improve the robustness of recording correct contact cases. As mentioned in Section~\ref{sec:RSATVM}, if a tuple in contact case is not verified due to the failure mentioned above, as long as there is one witness in $W$ verified the contact list $C$, the tuple will be considered valid. 
%A contact transaction $T_{con}$ is verified if all the pairs in $C$ or $W$ are verified. Therefore, even when $C$ is not fully verified, as long as $W$ is verified, which means we have witness contact case $C$, $C$ will be considered valid and be further stored with $T_{con}$ in blocks. 

To evaluate the robustness of recording contact information of the proposed system, we set a failure rate $p$ of each user, representing that the user has a probability $p$ of failing to verify the corresponding transaction. Then we compute how many contact cases, that $u_i$ contacts with $u_j$ at timestamp $t$, e.g. $(u_i, u_j, t)$, will lose comparing with the given 300k contact cases. We use a baseline $BDCT-w/o-Witness$ that is BDCT without witness role, therefore $(u_i, u_j, t)$ will be lost when both $u_i$ and $u_j$ fail to verify the corresponding tuples in $C$. This is also a common design in exiting works~\cite{DBLP:journals/access/AhmedMXRMKSHJJ20,DBLP:journals/health/ReichertBS21}. We simulate this experiment for 10 times, and report the average results.

\begin{table*}[htb!]
\centering
\caption{Average Contact Case Loss Percentage at Different Failure Rate $p$}
\label{tab:robust}
\begin{tabular}{ccccccc}
\hline
User failure rate  & p=0.1 & p=0.2 & p=0.3 & p=0.4 & p=0.5 & p=0.6 \\ \hline
 BDCT   & 0.18$\pm$0.02\% & 0.45$\pm$ 0.03\% & 0.68$\pm$0.01\%  & 1.19$\pm$0.02\%  & 1.96$\pm$0.12\%  & 3.69$\pm$0.03\%  \\
BDCT-w/o-Witness   & 2.25$\pm$0.03\% & 6.12$\pm$ 0.03\% & 11.12$\pm$ 0.04\% & 18.70$\pm$0.09\% & 27.63$\pm$0.07\% & 38.74$\pm$0.14\% \\ \hline
\end{tabular}
\end{table*}

Table~\ref{tab:robust} shows the simulation results. It can be seen that our framework lost significantly less contact cases than baseline at any failure rate $p$. BDCT can correctly record nearly 96.31\% (1-3.69\%) total contact cases even every node has 0.6 failure probability, which is 35\% more than the baseline that can only preserve about 61.26\% (1-38.74\%) contact cases.

\subsection{Attack Resistance}

Malicious users may report false or fake contact cases to generate more transactions, which may bring them more credit reward or stake reward. The only way to achieve this attack is through group cheating that several malicious users together create and verify transactions. More specifically, malicious users have two attack approaches to get a fake transaction verified. They can create a contact list where all the contacts are malicious users, or put malicious witness in the witness list so that the whole transaction will be verified as long as this malicious witness verifies related tuple in witness list no matter whether the contact case is real or fake in the contact list.  Next we describe how each of the attack approach impacts the whole system.  

If a malicious user chooses the first method, that creates contact list composed by other malicious users, this will not bring any false information to other honest users. The malicious users may earn more credits by generating or verifying numerous fake transactions, but the number of transactions is limited by the Bluetooth scanning frequency specified by the system, e.g. 5 minutes. In other words, a user can  generate at most twelve transactions per hour. In addition, the proposed RC-DPoS and incentive mechanism can well balance the stake reward as shown in Figure~\ref{fig:gini_results}. Therefore malicious users will not dominate stake reward too much. % Figure shows the Gini coefficient and the percentage of stake rewards received by all malicious user when there are 20\% malicious users forming a group in the system. 

If a malicious user attacks through the second method, that a malicious witness is put in the witness list for every false transaction, it may include false contact cases in the contact list. However, according to the proposed RSA-TVM method, the tuple recording the false contact case can be verified wrong and hence not be preserved before the transaction is put into the transaction pool. Therefore, the impact of this attack can also be controlled.

\subsection{Storage Cost Evaluation}
In BDCT framework, every user is holding the whole copy of blockchain where the contact transactions are stored. We evaluate the expected blockchain storage cost of the proposed BDCT framework by calculating the expected number of transactions and blocks generated per hour with respect to our experiment setting. 

We denote the expected size of total blockchain segment generated per hour as $E(S_{TB/h})$, expected size of all block heads per hour as $E(S_{BH/h})$, and expected size of all block bodies per hour as $E(S_{BB/h})$. The size of single block head is denoted as $s_{BH}$, the expected amount of blocks generated per hour is $N_{B/h}$. Hence, $E(S_{BH/h}) = E(N_{B/h})*s_{BH}$. The block bodies contains transactions, therefore $E(S_{BB/h}) = E(S_{T/h})$, where $E(S_{T/h})$ is the expected size of total transactions generated per hour. Consequently, $E(S_{TB/h})$ is calculated by Equation~\ref{eq:s_tb}. 
\begin{equation}
\label{eq:s_tb}
\begin{aligned}
    E(S_{TB/h}) &= E(S_{BH/h})+E(S_{BB/h})\\
        &=E(N_{B/h})*s_{BH} + E(S_{T/h})
\end{aligned}
\end{equation}

The speed of generating a block is predefined in the system, e.g. every 5 minutes. Therefore $E(N_{B/h})=12$. Based on the block structure illustrated in Figure~\ref{fig:blockstr}, $s_{BH}$ can be calculated as Equation~\ref{eq:s_bh}.
\begin{equation}
\label{eq:s_bh}
\begin{aligned}
s_{BH} &= s_{Block Hash}+s_{Previous Block Hash}\\
&+s_{u_{D_{ID}}}+s_{Timestamp}\\
&= 2*s_{Block Hash} +s_{u_{D_{ID}}} + s_{Timestamp}\\
& = 2*256\ bits + 10\ bytes + 32\ bits \\
& = 64\ bytes+ 10\ bytes + 4\ bytes = 78\ bytes
\end{aligned}
\end{equation}

$s_{Block Hash}$ is the size of a unique block ID, which is a SHA256 hash value, therefore $s_{Block Hash} = 256\ bits$. $s_{u_{D_{ID}}}$ is the size of block generator's device ID. In our framework, the set device ID a string contains 10 Hex characters, which can represent $2^{10*4} \approx 1.1\times10^{12}$ unique devices. Since each char type hex character take 1 byte,  $s_{u_{D_{ID}}} = 10\ bytes$. $s_{Timestamp}$ is size of an Unix timestamp of 32 bit integer type, hence $s_{Timestamp} = 32\ bits$.

In our experiment settings, three different contact scenarios are considered with different contact case generating frequency, number of contacted people and number of witnesses. Though there may also registration transactions in block bodies, registration transactions cost minor storage. In this discussion, we consider the general case that block bodies contain only contact transactions. Then $E(S_{T/h})$ is the sum of expected all transactions generated by 3 contact scenarios as in Equation~\ref{eq:s_th}. 
\begin{equation}
\label{eq:s_th}
\begin{aligned}
E(S_{T/h}) = E(S_{TS/h}) + E(S_{TM/h}) + E(S_{TC/h}) 
\end{aligned}
\end{equation}

$E(S_{TS/h})$, $E(S_{TM/h})$ and $E(S_{TC/h})$ are the expected numbers of transactions generated per hour in Sparse scenario, Medium scenario and Crowded scenario respectively. 

We denote $E(N_{TS/h})$ as the expected number of transactions (contact cases) per hour in Sparse scenario and  $E(s_{TS})$ as the expected size of a transaction generated in Sparse scenario. Given the contact transaction structure described in Section~\ref{sec:RSATVM}, $E(s_{TS})$ is composed by size of transaction ID $s_{T_{id}}$, size of transaction generator's device ID $s_{u_{D_{ID}}}$, size of transaction timestamp $s_{Timestamp}$, expected size of contact list $E(s_{C})$ and expected size of witness list $E(s_{W})$. In each contact list $C$ or witness list $W$, there are signed tuples $({u_i}_{Pub\_key}, SD_{{u_i}_{Pri\_key}})$. The size of signed tuples is denoted as $s_{st}$. In our experiments, we generate 1024 bits RSA Keys with the Python package Crypto\footnote{\url{https://pycryptodome.readthedocs.io}}. With the secret message as 10 Hex characters, $s_{st} = 56\ bytes+ 161\ bytes = 217\ bytes$ in our simulation.

Though the length of contact list and witness list are sampled from normal distribution described in Section~\ref{sec:sim_method}, we set the length to 0 if the sampled length is less than 0. The expected length of contact list $E(N_{C})$ based such sample strategy satisfies Equation~\ref{eq:nc}.

\begin{equation}
\label{eq:nc}
\begin{aligned}
E(N_{C}) &= \int_{0}^{\infty} x \frac{1}{\sigma \sqrt{2\pi}} \exp(-\frac{(x-\mu)^2}{2\sigma^2})\, dx \\
&= \frac{1}{\sigma \sqrt{2\pi}} \int_{0}^{\infty} x \exp(-\frac{(x-\mu)^2}{2\sigma^2})\, dx \\
&= \frac{\sqrt{2}\sigma}{\sigma \sqrt{2\pi}}\int_{-\frac{\mu}{\sqrt{2}\sigma}}^{\infty}(\sqrt{2}\sigma t+\mu)\exp^{-t^2}\, dt (with\ t= \frac{x-\mu}{\sqrt{2}\sigma})\\
&= \frac{1}{\sqrt{\pi}}(\sqrt{2}\sigma\int_{-\frac{\mu}{\sqrt{2}\sigma}}^{\infty} t\exp^{-t^2}\, dt + \mu \int_{-\frac{\mu}{\sqrt{2}\sigma}}^{\infty}\exp^{-t^2}\, dt)\\
&= \frac{1}{\sqrt{\pi}}(\sqrt{2}\sigma \left[-\frac{1}{2}\exp^{-t^2}\right]_{-\frac{\mu}{\sqrt{2}\sigma}}^{\infty} + \mu \int_{-\frac{\mu}{\sqrt{2}\sigma}}^{\infty}\exp^{-t^2}\, dt))\\
&= \frac{1}{\sqrt{\pi}} (\frac{\sigma}{\sqrt{2}} \exp^{-\frac{\mu^2}{2\sigma^2}} + \mu Z(t))
\end{aligned}
\end{equation}

\begin{equation}
\label{eq:zt}
\begin{aligned}
Z(t) &= \int_{-\frac{\mu}{\sqrt{2}\sigma}}^{\infty}\exp^{-t^2}\, dt\\
    &= \int_{-\frac{\mu}{\sqrt{2}\sigma}}^{0}\exp^{-t^2}\, dt + \int_{0}^{\infty}\exp^{-t^2}\, dt \\
    &= \int_{0}^{\frac{\mu}{\sqrt{2}\sigma}}\exp^{-t^2}\, dt + \frac{1}{2} \int_{-\infty}^{\infty}\exp^{-t^2}\, dt \\
    &= \int_{0}^{\frac{\mu}{\sqrt{2}\sigma}}\exp^{-t^2}\, dt + \frac{\sqrt{\pi}}{2}\\
    &\leq \int_{0}^{\frac{\mu}{\sqrt{2}\sigma}}\frac{1}{t^2+1}\, dt + \frac{\sqrt{\pi}}{2}\\
    &= arctan(\frac{\mu}{\sqrt{2}\sigma}) + \frac{\sqrt{\pi}}{2}
\end{aligned}
\end{equation}

\begin{equation}
\label{eq:s_tsh}
\begin{aligned}
E(S_{TS/h}) &= E(N_{TS/h})*E(s_{TS}) \\
            &= 1*(s_{T_{id}}+s_{u_{D_{ID}}}+s_{Timestamp} \\
            &+ E(s_{C}) +E(s_{W}) )\\
            &= 256\ bits + 10\ bytes + 32\ bits +  E(s_{C}) +E(s_{W}) \\
            &= 46\ bytes + E(N_{C})*s_{st} +  E(N_{W})*s_{st} \\ 
            &\leq 46\ bytes + (\frac{\sqrt{2}}{\sqrt{\pi}} + \frac{\sqrt{2}}{2\sqrt{\pi}})*217\ bytes\\
            &\approx 306\ bytes.
\end{aligned}
\end{equation}

Since $Z(t)$ is limited by an upper bond, we can calculate upper bond of the expected number of transactions per hour in Sparse scenario as Equation~\ref{eq:s_tsh}. Similarly, $E(S_{TM/h}) \leq 3374\ bytes$ and $E(S_{TC/h})\leq 36227\ bytes$.

Hence $E(S_{T/h})\leq 306+3374+36227 = 39907\ bytes$. Then $E(S_{TB/h})=E(N_{B/h})*s_{BH} + E(S_{T/h}) = 12*78+39907\ bytes= 40843\ bytes \approx 39.89\ KB$. Therefore the expected storage cost for the blockchain generated per day is $24*E(S_{T/h}) = 24*39.89\ KB = 957.36\ KB < 1 MB$, which is totally affordable for most smart devices. 

% \begin{equation}
% \label{eq:s_tmh}
% \begin{aligned}
% E(S_{TM/h}) = E(N_{TM/h})*E(s_{TM}) 
% \end{aligned}
% \end{equation}

% \begin{equation}
% \label{eq:s_tch}
% \begin{aligned}
% E(S_{TC/h}) = E(N_{TC/h})*E(s_{CS}) 
% \end{aligned}
% \end{equation}

\section{Conclusion}
\label{sec:con}
In this article, we propose a Blockchain Driven Contact Tracing framework (BDCT), which is a fully decentralized framework without any third-party required. We propose the role ``witness" in the framework to promote contact tracing data integrity, and the RSA based Transaction Verification Method (RSA-TVM) to verify the correctness of the reported contact cases. Reputation Corrected Delegated Proof of Stake (RC-DPoS) consensus mechanism is applied to select miners based on both users' reputation and users' stake. An incentive mechanism is further developed to motivate people to keep reporting contact cases honestly and work with RC-DPoS achieving balanced stake reward distribution to maintain the whole framework decentralized. 
In the simulation, we propose a simulation environment, which mixes three contact scenarios based on different population density. The simulation results demonstrate our proposed framework can achieve significantly decentrality than the baseline framework, and RSA-TVM incorporated with ``witness" role in the framework can hugely improve the system robustness.

For future work, better consensus mechanism should be designed to lower communication cost. In BDCT, though the computation cost is eliminated and storage cost is considered acceptable, the communication cost is still high for circulating contact transactions among devices to get contact cases verified especially when contact cases are tremendous. Synchronizing blockchain and shared transaction pool also imposes communication stress on smart devices. Therefore a better communication protocol is highly demanded for making more scalable contact tracing applications.

%For future work, we will deeply investigate better blockchain storage methods. Since each user needs to store the whole copy of the blockchain, the storage cost will gradually become unacceptable for mobile devices with the growth of blockchain when the user number is huge. Though it is believed contact history of 14 days is enough for detecting potential infected cases, recording 14-days contact information may still be a challenge for mobile devices when contact cases are tremendous. Therefore a better blockchain storage method is highly demanded for making more scalable contact tracing applications. 

% or



\ifCLASSOPTIONcaptionsoff
  \newpage
\fi



\bibliographystyle{IEEEtran}
\bibliography{ref}

\begin{thebibliography}{10}
\providecommand{\url}[1]{#1}
\csname url@samestyle\endcsname
\providecommand{\newblock}{\relax}
\providecommand{\bibinfo}[2]{#2}
\providecommand{\BIBentrySTDinterwordspacing}{\spaceskip=0pt\relax}
\providecommand{\BIBentryALTinterwordstretchfactor}{4}
\providecommand{\BIBentryALTinterwordspacing}{\spaceskip=\fontdimen2\font plus
\BIBentryALTinterwordstretchfactor\fontdimen3\font minus
  \fontdimen4\font\relax}
\providecommand{\BIBforeignlanguage}[2]{{%
\expandafter\ifx\csname l@#1\endcsname\relax
\typeout{** WARNING: IEEEtran.bst: No hyphenation pattern has been}%
\typeout{** loaded for the language `#1'. Using the pattern for}%
\typeout{** the default language instead.}%
\else
\language=\csname l@#1\endcsname
\fi
#2}}
\providecommand{\BIBdecl}{\relax}
\BIBdecl

\bibitem{Dong2020AnIW}
E.~Dong, H.~Du, and L.~Gardner, ``An interactive web-based dashboard to track
  covid-19 in real time,'' \emph{The Lancet. Infectious Diseases}, vol.~20, pp.
  533 -- 534, 2020.

\bibitem{world2014contact}
W.~H. Organization \emph{et~al.}, ``Contact tracing during an outbreak of ebola
  virus disease,'' 2014.

\bibitem{Bay2020BlueTraceAP}
J.~Bay, J.~Kek, A.~Tan, and C.~S. Hau, ``Bluetrace: A privacy-preserving
  protocol for community-driven contact tracing across borders,'' 2020.

\bibitem{Liang2020}
\BIBentryALTinterwordspacing
F.~Liang, ``Covid-19 and health code: How digital platforms tackle the pandemic
  in china,'' \emph{Social Media + Society}, vol.~6, no.~3, p.
  2056305120947657, 2020, pMID: 34192023. [Online]. Available:
  \url{https://doi.org/10.1177/2056305120947657}
\BIBentrySTDinterwordspacing

\bibitem{apple_newsroom_2021}
\BIBentryALTinterwordspacing
Apple and Google, ``Apple and google partner on covid-19 contact tracing
  technology,'' Dec 2021. [Online]. Available:
  \url{https://www.apple.com/newsroom/2020/04/apple-and-google-partner-on-covid-19-contact-tracing-technology/}
\BIBentrySTDinterwordspacing

\bibitem{hasan2021covid}
H.~R. Hasan, K.~Salah, R.~Jayaraman, I.~Yaqoob, M.~Omar, and S.~Ellahham,
  ``Covid-19 contact tracing using blockchain,'' \emph{IEEE Access}, 2021.

\bibitem{DBLP:journals/iotj/XuZOFBI21}
\BIBentryALTinterwordspacing
H.~Xu, L.~Zhang, O.~Onireti, Y.~Fang, W.~J. Buchanan, and M.~A. Imran,
  ``Beeptrace: Blockchain-enabled privacy-preserving contact tracing for
  {COVID-19} pandemic and beyond,'' \emph{{IEEE} Internet Things J.}, vol.~8,
  no.~5, pp. 3915--3929, 2021. [Online]. Available:
  \url{https://doi.org/10.1109/JIOT.2020.3025953}
\BIBentrySTDinterwordspacing

\bibitem{DBLP:journals/tnse/LvWJCQZ22}
\BIBentryALTinterwordspacing
W.~Lv, S.~Wu, C.~Jiang, Y.~Cui, X.~Qiu, and Y.~Zhang, ``Towards large-scale and
  privacy-preserving contact tracing in {COVID-19} pandemic: {A} blockchain
  perspective,'' \emph{{IEEE} Trans. Netw. Sci. Eng.}, vol.~9, no.~1, pp.
  282--298, 2022. [Online]. Available:
  \url{https://doi.org/10.1109/TNSE.2020.3030925}
\BIBentrySTDinterwordspacing

\bibitem{Altmann2020AcceptabilityOA}
S.~Altmann, L.~Milsom, H.~Zillessen, R.~Blasone, F.~Gerdon, R.~L. Bach,
  F.~Kreuter, D.~Nosenzo, S.~Toussaert, and J.~Abeler, ``Acceptability of
  app-based contact tracing for covid-19: Cross-country survey study,''
  \emph{JMIR mHealth and uHealth}, vol.~8, 2020.

\bibitem{Altuwaiyan2018EPICEP}
T.~Altuwaiyan, M.~Hadian, and X.~Liang, ``Epic: Efficient privacy-preserving
  contact tracing for infection detection,'' \emph{2018 IEEE International
  Conference on Communications (ICC)}, pp. 1--6, 2018.

\bibitem{Prasad2017ENACTEA}
A.~Prasad and D.~F. Kotz, ``Enact: Encounter-based architecture for contact
  tracing,'' \emph{Proceedings of the 4th International on Workshop on Physical
  Analytics}, 2017.

\bibitem{nisar2020privacy}
S.~Nisar, M.~A. Zuhaib, A.~Ulasyar, and M.~Tariq, ``A privacy preserved and
  cost efficient control scheme for coronavirus outbreak using call data record
  and contact tracing,'' \emph{IEEE Consumer Electronics Magazine}, 2020.

\bibitem{Trivedi2021WiFiTraceNC}
A.~Trivedi, C.~Zakaria, R.~K. Balan, and P.~J. Shenoy, ``Wifitrace:
  Network-based contact tracing for infectious diseases using passive wifi
  sensing,'' \emph{Proc. ACM Interact. Mob. Wearable Ubiquitous Technol.},
  vol.~5, pp. 37:1--37:26, 2021.

\bibitem{Hekmati2021CONTAINPC}
A.~Hekmati, G.~S. Ramachandran, and B.~Krishnamachari, ``Contain:
  Privacy-oriented contact tracing protocols for epidemics,'' \emph{2021
  IFIP/IEEE International Symposium on Integrated Network Management (IM)}, pp.
  872--877, 2021.

\bibitem{Ng2021COVID19AY}
P.~C. Ng, P.~Spachos, and K.~N. Plataniotis, ``Covid-19 and your smartphone:
  Ble-based smart contact tracing,'' \emph{IEEE Systems Journal}, vol.~15, pp.
  5367--5378, 2021.

\bibitem{reichert2020privacy}
L.~Reichert, S.~Brack, and B.~Scheuermann, ``Privacy-preserving contact tracing
  of covid-19 patients.'' \emph{IACR Cryptol. ePrint Arch.}, vol. 2020, p. 375,
  2020.

\bibitem{chan2020pact}
J.~Chan, D.~Foster, S.~Gollakota, E.~Horvitz, J.~Jaeger, S.~Kakade, T.~Kohno,
  J.~Langford, J.~Larson, P.~Sharma \emph{et~al.}, ``Pact: Privacy sensitive
  protocols and mechanisms for mobile contact tracing,'' \emph{arXiv preprint
  arXiv:2004.03544}, 2020.

\bibitem{DBLP:journals/iotj/AzadAARAIA21}
\BIBentryALTinterwordspacing
M.~A. Azad, J.~Arshad, S.~M.~A. Akmal, F.~Riaz, S.~Abdullah, M.~Imran, and
  F.~Ahmad, ``A first look at privacy analysis of {COVID-19} contact-tracing
  mobile applications,'' \emph{{IEEE} Internet Things J.}, vol.~8, no.~21, pp.
  15\,796--15\,806, 2021. [Online]. Available:
  \url{https://doi.org/10.1109/JIOT.2020.3024180}
\BIBentrySTDinterwordspacing

\bibitem{DBLP:journals/access/AhmedMXRMKSHJJ20}
\BIBentryALTinterwordspacing
N.~Ahmed, R.~A. Michelin, W.~Xue, S.~Ruj, R.~A. Malaney, S.~S. Kanhere,
  A.~Seneviratne, W.~Hu, H.~Janicke, and S.~K. Jha, ``A survey of {COVID-19}
  contact tracing apps,'' \emph{{IEEE} Access}, vol.~8, pp. 134\,577--134\,601,
  2020. [Online]. Available: \url{https://doi.org/10.1109/ACCESS.2020.3010226}
\BIBentrySTDinterwordspacing

\bibitem{Nakamoto2009BitcoinA}
S.~Nakamoto, ``Bitcoin : A peer-to-peer electronic cash system,'' 2009.

\bibitem{DBLP:journals/iotj/GuoDW21}
\BIBentryALTinterwordspacing
J.~Guo, X.~Ding, and W.~Wu, ``A blockchain-enabled ecosystem for distributed
  electricity trading in smart city,'' \emph{{IEEE} Internet Things J.},
  vol.~8, no.~3, pp. 2040--2050, 2021. [Online]. Available:
  \url{https://doi.org/10.1109/JIOT.2020.3015980}
\BIBentrySTDinterwordspacing

\bibitem{DBLP:journals/tcss/FanWWD21}
\BIBentryALTinterwordspacing
Y.~Fan, L.~Wang, W.~Wu, and D.~Du, ``Cloud/edge computing resource allocation
  and pricing for mobile blockchain: An iterative greedy and search approach,''
  \emph{{IEEE} Trans. Comput. Soc. Syst.}, vol.~8, no.~2, pp. 451--463, 2021.
  [Online]. Available: \url{https://doi.org/10.1109/TCSS.2021.3049152}
\BIBentrySTDinterwordspacing

\bibitem{DBLP:journals/tnse/DingGLW21}
\BIBentryALTinterwordspacing
X.~Ding, J.~Guo, D.~Li, and W.~Wu, ``An incentive mechanism for building a
  secure blockchain-based internet of things,'' \emph{{IEEE} Trans. Netw. Sci.
  Eng.}, vol.~8, no.~1, pp. 477--487, 2021. [Online]. Available:
  \url{https://doi.org/10.1109/TNSE.2020.3040446}
\BIBentrySTDinterwordspacing

\bibitem{DBLP:journals/tr/DongWGSZD21}
\BIBentryALTinterwordspacing
L.~Dong, W.~Wu, Q.~Guo, M.~N. Satpute, T.~Znati, and D.~Du, ``Reliability-aware
  offloading and allocation in multilevel edge computing system,'' \emph{{IEEE}
  Trans. Reliab.}, vol.~70, no.~1, pp. 200--211, 2021. [Online]. Available:
  \url{https://doi.org/10.1109/TR.2019.2909279}
\BIBentrySTDinterwordspacing

\bibitem{DBLP:conf/birthday/LuoX0W20}
\BIBentryALTinterwordspacing
C.~Luo, L.~Xu, D.~Li, and W.~Wu, ``Edge computing integrated with blockchain
  technologies,'' in \emph{Complexity and Approximation - In Memory of Ker-I
  Ko}, ser. Lecture Notes in Computer Science, D.~Du and J.~Wang, Eds., vol.
  12000.\hskip 1em plus 0.5em minus 0.4em\relax Springer, 2020, pp. 268--288.
  [Online]. Available: \url{https://doi.org/10.1007/978-3-030-41672-0\_17}
\BIBentrySTDinterwordspacing

\bibitem{idrees2021blockchain}
S.~M. Idrees, M.~Nowostawski, and R.~Jameel, ``Blockchain-based digital contact
  tracing apps for covid-19 pandemic management: Issues, challenges, solutions,
  and future directions,'' \emph{JMIR medical informatics}, vol.~9, no.~2, p.
  e25245, 2021.

\bibitem{arifeen2020blockchain}
M.~M. Arifeen, A.~Al~Mamun, M.~S. Kaiser, and M.~Mahmud, ``Blockchain-enable
  contact tracing for preserving user privacy during covid-19 outbreak,'' 2020.

\bibitem{DBLP:journals/csi/ZhangXSZ21}
\BIBentryALTinterwordspacing
C.~Zhang, C.~Xu, K.~Sharif, and L.~Zhu, ``Privacy-preserving contact tracing in
  5g-integrated and blockchain-based medical applications,'' \emph{Comput.
  Stand. Interfaces}, vol.~77, p. 103520, 2021. [Online]. Available:
  \url{https://doi.org/10.1016/j.csi.2021.103520}
\BIBentrySTDinterwordspacing

\bibitem{DBLP:conf/sigmod/PengXWHXC21}
\BIBentryALTinterwordspacing
Z.~Peng, C.~Xu, H.~Wang, J.~Huang, J.~Xu, and X.~Chu,
  ``P\({}^{\mbox{2}}\)b-trace: Privacy-preserving blockchain-based contact
  tracing to combat pandemics,'' in \emph{{SIGMOD} '21: International
  Conference on Management of Data, Virtual Event, China, June 20-25, 2021},
  G.~Li, Z.~Li, S.~Idreos, and D.~Srivastava, Eds.\hskip 1em plus 0.5em minus
  0.4em\relax {ACM}, 2021, pp. 2389--2393. [Online]. Available:
  \url{https://doi.org/10.1145/3448016.3459237}
\BIBentrySTDinterwordspacing

\bibitem{DBLP:journals/sncs/VangipuramMK21}
\BIBentryALTinterwordspacing
S.~L.~T. Vangipuram, S.~P. Mohanty, and E.~Kougianos, ``Covichain: {A}
  blockchain based framework for nonrepudiable contact tracing in healthcare
  cyber-physical systems during pandemic outbreaks,'' \emph{{SN} Comput. Sci.},
  vol.~2, no.~5, p. 346, 2021. [Online]. Available:
  \url{https://doi.org/10.1007/s42979-021-00746-x}
\BIBentrySTDinterwordspacing

\bibitem{zuhair2021blocov6}
M.~Zuhair, F.~Patel, D.~Navapara, P.~Bhattacharya, and D.~Saraswat, ``Blocov6:
  A blockchain-based 6g-assisted uav contact tracing scheme for covid-19
  pandemic,'' in \emph{2021 2nd International Conference on Intelligent
  Engineering and Management (ICIEM)}.\hskip 1em plus 0.5em minus 0.4em\relax
  IEEE, 2021, pp. 271--276.

\bibitem{DBLP:journals/corr/abs-2108-08275}
\BIBentryALTinterwordspacing
M.~Salimibeni, Z.~Hajiakhondi{-}Meybodi, A.~Mohammadi, and Y.~Wang, ``{TB-ICT:}
  {A} trustworthy blockchain-enabled system for indoor {COVID-19} contact
  tracing,'' \emph{CoRR}, vol. abs/2108.08275, 2021. [Online]. Available:
  \url{https://arxiv.org/abs/2108.08275}
\BIBentrySTDinterwordspacing

\bibitem{kotanen2003experiments}
A.~Kotanen, M.~Hannikainen, H.~Leppakoski, and T.~D. Hamalainen, ``Experiments
  on local positioning with bluetooth,'' in \emph{Proceedings ITCC 2003.
  International Conference on Information Technology: Coding and
  Computing}.\hskip 1em plus 0.5em minus 0.4em\relax IEEE, 2003, pp. 297--303.

\bibitem{rivest1978method}
R.~L. Rivest, A.~Shamir, and L.~Adleman, ``A method for obtaining digital
  signatures and public-key cryptosystems,'' \emph{Communications of the ACM},
  vol.~21, no.~2, pp. 120--126, 1978.

\bibitem{DBLP:conf/sacrypt/GilbertH03}
\BIBentryALTinterwordspacing
H.~Gilbert and H.~Handschuh, ``Security analysis of {SHA-256} and sisters,'' in
  \emph{Selected Areas in Cryptography, 10th Annual International Workshop,
  {SAC} 2003, Ottawa, Canada, August 14-15, 2003, Revised Papers}, ser. Lecture
  Notes in Computer Science, M.~Matsui and R.~J. Zuccherato, Eds., vol.
  3006.\hskip 1em plus 0.5em minus 0.4em\relax Springer, 2003, pp. 175--193.
  [Online]. Available: \url{https://doi.org/10.1007/978-3-540-24654-1\_13}
\BIBentrySTDinterwordspacing

\bibitem{larimer2014delegated}
D.~Larimer, ``Delegated proof-of-stake (dpos),'' \emph{Bitshare whitepaper},
  vol.~81, p.~85, 2014.

\bibitem{DBLP:journals/simpra/JahromiZG016}
\BIBentryALTinterwordspacing
K.~K. Jahromi, M.~Zignani, S.~Gaito, and G.~P. Rossi, ``Simulating human
  mobility patterns in urban areas,'' \emph{Simul. Model. Pract. Theory},
  vol.~62, pp. 137--156, 2016. [Online]. Available:
  \url{https://doi.org/10.1016/j.simpat.2015.12.002}
\BIBentrySTDinterwordspacing

\bibitem{huang2019grab}
X.~Huang, Y.~Yin, S.~Lim, G.~Wang, B.~Hu, J.~Varadarajan, S.~Zheng, A.~Bulusu,
  and R.~Zimmermann, ``Grab-posisi: An extensive real-life gps trajectory
  dataset in southeast asia,'' in \emph{Proceedings of the 3rd ACM SIGSPATIAL
  International Workshop on Prediction of Human Mobility}, 2019, pp. 1--10.

\bibitem{didi}
\BIBentryALTinterwordspacing
C.~DiDi, ``Data source: Didi chuxing gaia open dataset initiative,'' 2019.
  [Online]. Available:
  \url{https://outreach.didichuxing.com/research/opendata/en/}
\BIBentrySTDinterwordspacing

\bibitem{DBLP:journals/tsas/Mariescu-Istodor18}
\BIBentryALTinterwordspacing
R.~Mariescu{-}Istodor and P.~Fr{\"{a}}nti, ``Cellnet: Inferring road networks
  from {GPS} trajectories,'' \emph{{ACM} Trans. Spatial Algorithms Syst.},
  vol.~4, no.~3, pp. 8:1--8:22, 2018. [Online]. Available:
  \url{https://doi.org/10.1145/3234692}
\BIBentrySTDinterwordspacing

\bibitem{DBLP:conf/gis/YuanZZXXSH10}
\BIBentryALTinterwordspacing
J.~Yuan, Y.~Zheng, C.~Zhang, W.~Xie, X.~Xie, G.~Sun, and Y.~Huang, ``T-drive:
  driving directions based on taxi trajectories,'' in \emph{18th {ACM}
  {SIGSPATIAL} International Symposium on Advances in Geographic Information
  Systems, {ACM-GIS} 2010, November 3-5, 2010, San Jose, CA, USA, Proceedings},
  D.~Agrawal, P.~Zhang, A.~E. Abbadi, and M.~F. Mokbel, Eds.\hskip 1em plus
  0.5em minus 0.4em\relax {ACM}, 2010, pp. 99--108. [Online]. Available:
  \url{https://doi.org/10.1145/1869790.1869807}
\BIBentrySTDinterwordspacing

\bibitem{DBLP:conf/sensys/LianZ18}
\BIBentryALTinterwordspacing
J.~Lian and L.~Zhang, ``One-month beijing taxi {GPS} trajectory dataset with
  taxi ids and vehicle status,'' in \emph{Proceedings of the First Workshop on
  Data Acquisition To Analysis, DATA@SenSys 2018, Shenzhen, China, November 4,
  2018}, J.~Gao, P.~Zhang, S.~Pan, and C.~Ni, Eds.\hskip 1em plus 0.5em minus
  0.4em\relax {ACM}, 2018, pp. 3--4. [Online]. Available:
  \url{https://doi.org/10.1145/3277868.3277870}
\BIBentrySTDinterwordspacing

\bibitem{kakwani1977}
\BIBentryALTinterwordspacing
N.~C. Kakwani, ``Applications of lorenz curves in economic analysis,''
  \emph{Econometrica}, vol.~45, no.~3, pp. 719--727, 1977. [Online]. Available:
  \url{http://www.jstor.org/stable/1911684}
\BIBentrySTDinterwordspacing

\bibitem{dorfman1979formula}
R.~Dorfman, ``A formula for the gini coefficient,'' \emph{The review of
  economics and statistics}, pp. 146--149, 1979.

\bibitem{DBLP:journals/health/ReichertBS21}
\BIBentryALTinterwordspacing
L.~Reichert, S.~Brack, and B.~Scheuermann, ``A survey of automatic contact
  tracing approaches using bluetooth low energy,'' \emph{{ACM} Trans. Comput.
  Heal.}, vol.~2, no.~2, pp. 18:1--18:33, 2021. [Online]. Available:
  \url{https://doi.org/10.1145/3444847}
\BIBentrySTDinterwordspacing

\end{thebibliography}


\begin{IEEEbiography}[{\includegraphics[width=1in,height=1.25in,clip,keepaspectratio]{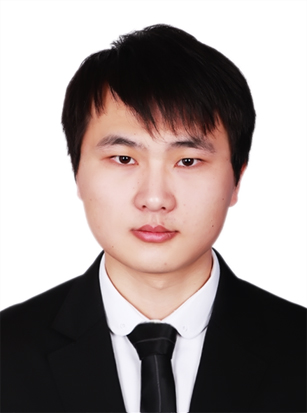}}]{Xiao Li}
received his B.S. and M.S degree in Software Engineering from
Dalian University of Technology, China in 2016 and 2019, respectively. He is
currently pursuing the Ph.D. degree with the Department of Computer Science,
University of Texas at Dallas, Richardson, TX, USA. His current research
interests include data mining and Blockchain.
\end{IEEEbiography}

\begin{IEEEbiography}[{\includegraphics[width=1in,height=1.25in,clip,keepaspectratio]{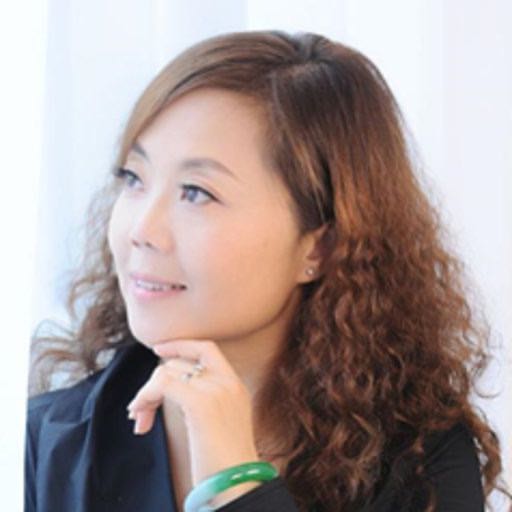}}]{Weili Wu}
(Senior Member, IEEE) received the M.S.
and Ph.D. degrees from the Department of Computer
Science, University of Minnesota, Minneapolis, MN,
USA, in 1998 and 2002, respectively.
She is currently a Full Professor with the
Department of Computer Science, University of
Texas at Dallas, Richardson, TX, USA. Her research
mainly deals in the general research area of data
communication and data management. Her research
focuses on the design and analysis of algorithms
for optimization problems that occur in wireless
networking environments and various database systems.
\end{IEEEbiography}

%\newpage

\begin{IEEEbiography}[{\includegraphics[width=1in,height=1.25in,clip,keepaspectratio]{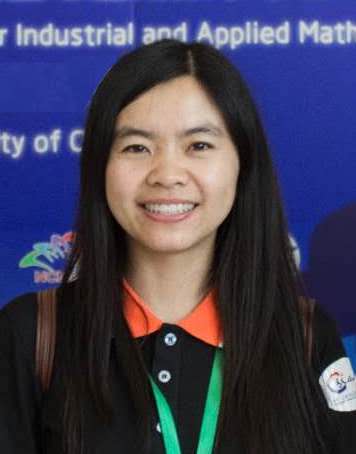}}]{Tiantian Chen}
 is a Ph.D. candidate in the Department of Computer Science, The University of Texas at Dallas. She received her B.S. degree in Mathematics and Applied Mathematics, and M.S. degree in Operational Research and Cybernetics from Ocean University of China in 2016 and 2019, respectively. Her research focuses on social networks, design and analysis of approximation algorithms, deep learning, and reinforcement learning.
\end{IEEEbiography}


%\vfill


\end{document}